\documentclass[12pt]{article}
\usepackage{amsmath,amsfonts,epsfig}
\numberwithin{equation}{section}

\ifx\ltimes\undefined
\newcommand{\ltimes}{{\kern3pt\hbox{\vrule width 0.4pt height 5.30pt
depth .0pt}\kern-1.76pt\times\kern1pt}} \fi

\def\Z {\mathbb{Z}}
\def\R {\mathbb{R}}

\def\a{\alpha}
\def\b{\beta}

\def\m{\mu}

\def\si{\sigma}                                   %     \varsigma

\def\G{\Gamma}

\def\cX{{\cal X}}

\begin{document}

\begin{titlepage}
\begin{flushleft}
\hfill  hep-th/0610263 \\
\hfill  Imperial/TP/06/RAR/04 \\

\end{flushleft}
\vspace*{8mm}

\begin{center}

{\Large {Geometric and Non-Geometric\\ Compactifications of IIB Supergravity}} \\

\vspace*{12mm}

{ R.A.~Reid-Edwards
} \\
\vspace*{7mm}

{\em The Institute for Mathematical Sciences} \\ {\em Imperial College of Science and Technology} \\
{\em 53 Prince's Gate, London SW7 2PG, UK} \\

\vspace*{12mm}

\end{center}

\begin{abstract}
Complimentary geometric and non-geometric consistent reductions of IIB supergravity are studied. The geometric reductions on the identified
group manifold $\cX=G/\G$ are found to have a gauge symmetry with Lie algebroid structure, generalising that found in similar reductions of the
Bosonic string theory and eleven-dimensional supergravity. Examples of such compactifications are considered and the symmetry breaking in each
case is analysed. Complimentary to the reductions on $\cX$ are the nine-dimensional S-duality twisted reductions considered in the second half
of the paper. The general reduced theory is given and symmetry breaking is investigated. The non-geometric S-duality twisted reductions and
their relation to geometric reductions of F-Theory on $\cX$ is briefly discussed.
\end{abstract}

\vfill

\noindent {Email: { r.reid-edwards@imperial.ac.uk} }

\end{titlepage}

\newpage

\section{Introduction}

The majority of techniques used to study the dimensional reduction of $D+d$ dimensional Supergravity theories to $D$ space-time dimensions may
be divided into two broad categories.

The first, and most amenable to physical interpretation, requires the higher-dimensional theory to be rewritten in terms of the harmonic modes
of some compact internal manifold and gives rise to an infinite tower of massive Kaluza-Klein fields corresponding to excitations of the higher
harmonics. This results in a rather cumbersome rewriting of the original theory and in practice only a truncation of the theory to the lightest
modes is considered \cite{Salam:1981xd}. The drawbacks of this approach are that the treatment of the harmonic modes can be laborious in
practice and the low energy modes that remain in the truncated theory generally will not solve the higher-dimensional equations of motion. Thus
solutions of the truncated theory generally will not lift to solutions of the higher dimensional theory. This technique will be referred to as
Kaluza-Klein reduction. The mass of the modes in the Kaluza-Klein reduction are given by the eigenvalues of the Laplacian on the internal
manifold. As such the truncation to the lowest modes keeps the lightest states whose number is given by the Betti numbers of the internal
manifold

The second approach, and the one that will be considered in this paper, is to consider solutions of the higher dimensional equations of motion
that may be given a lower dimensional \emph{interpretation}. This lower dimensional interpretation arises from the requirement that the solution
does not depend on the internal coordinates and generally will not involve all of the fields in the higher dimensional theory. Such a truncation
of the spectrum, that solves the higher dimensional equations of motion, is called \emph{consistent} although the term is now generally applied
to any solution of the full theory that is independent of the internal coordinates, regardless of whether or not the solution arose from a
truncation of a harmonic expansion. In contrast to the Kaluza-Klein approach the effective theory obtained may not include all of the lightest
modes and the number of preserved fields will not generally be related to the topological properties of the internal manifold.

There is some overlap between these categories and examples exist of Kaluza-Klein truncations which are consistent. An interesting example is
compactification on a torus for which a truncation to the lightest harmonics (Fourier modes) yields an effective lower dimensional theory. The
resulting effective theory solves the higher dimensional equations of motion and is an example of both procedures. As a counter-example,
compactification on a Calabi-Yau followed by truncation to the zero modes is not consistent in this sense.

The consistency of a reduction can often be understood group theoretically\footnote{Coset reductions appear to be an exception. There is, as
yet, no systematic understanding of the consistency of such reductions, despite the numerous examples that exist
\cite{Duff:1986hr,Cvetic:2003jy,Cvetic:2000dm}.}. For example, in the case of a reduction on a torus discussed above, the momenta of the
compactified fields along the internal directions give rise to conserved charges in the $D$-dimensional theory. Truncating to the zero modes in
this case leads to keeping only the singlets, therefore there can be no chance of interactions generating the modes that have been truncated
out. This truncation is therefore consistent.

Various generalisations of the toroidal reduction which are consistent (i.e. do not depend on the internal coordinates and solve the
higher-dimensional equations of motion) have been systematically studied
\cite{Duff:1986hr,Cvetic:2003jy,Cvetic:2000dm,Lavrinenko:1996mp,Lavrinenko:1997qa,Scherk:1978ta,Scherk:1979zr}. These include the Scherk-Schwarz
reductions \cite{Scherk:1978ta,Scherk:1979zr} and reductions with cohomological fluxes \cite{Lavrinenko:1996mp,Lavrinenko:1997qa} which yield
effective theories with scalar potentials and non-abelian gauge symmetries. These reductions may be thought of as massive deformations of the
toroidal solution.

In this paper we consider Scherk-Schwarz reductions of IIB supergravity and F-Theory that admit a geometric interpretation\footnote{We describe
the Scherk-Schwarz solution as a compactification on $\cX$ in the sense that the reduction ansatz is constructed from the set of globally
defined one-forms on $\cX$.} as a compactification on a manifold $\cX=G/\G$. Here $G$ is a, possibly non-compact, $d$-dimensional group manifold
and $\G\subset G$ is a discrete subgroup acting from the left such that $\cX$ is compact. The group manifold $G$ has isometry group $G_L\times
G_R$ (the action of the group on itself from the left and right respectively) which is broken to $G_R$ by the discrete quotient (see
\cite{Hull:2005hk} for a detailed discussion), leaving an effective theory with $G_R$ gauge symmetry. In order for the reduced fields to be
globally defined on $\cX$ the Scherk-Schwarz ansatz requires the reduction ansatz to be invariant under the rigid action of $G_L$
\cite{Hull:2005hk,Hull:2006tp}. The spaces $\cX$ are known as twisted tori in the literature. Generally $\cX$ will bare no relation to a torus
fibration so the terminology is misleading. Henceforth $\cX=G/\G$ shall be referred to as an \emph{identified group manifold}\footnote{Suggested
by C. Hull. Other names one might consider are `Cocompact Orbifold' or `Coset', but these names are misleading as $\G$ is discrete and acts
freely on $G$.}. Examples of identified-group manifolds that are topologically twisted torus fibrations were given in \cite{Hull:2005hk}.

This construction and its generalisation to include matter with flux is reviewed in the following section. Section 3 presents a study of the
flux compactification of IIB Supergravity on identified group manifolds with particular focus on the gauge symmetry and its breaking. The
Kaluza-Klein reduction has a clear geometric interpretation by construction. The consistent reductions may not always be easily identified with
the truncation of a Kaluza-Klein reduction on some manifold and it is interesting to consider the higher-dimensional origin of these reductions.
In contrast to the reduction via harmonic analysis, many solutions generally cannot be interpreted in terms of a geometric compactification.
Many examples now exist \cite{Hull:2005hk,Hull:2006tp,Hull:2003kr,Catal-Ozer:2006mn,Hull:2006va,Hull:2006qs,Dabholkar:2002sy,Dabholkar:2005ve}
of solutions of the higher dimensional theory, depending only a set of macroscopic coordinates, for which the internal space cannot be
understood in terms of classical Riemannian geometry. In section 4 we consider S-duality twisted reductions of IIB Supergravity. These
reductions do not arise from a geometric reduction of IIB Supergravity but may be interpreted as a compactification of F-Theory on an identified
group manifold.

\section{Flux Compactifications on Identified Group Manifolds}

In this section the Scherk-Schwarz reduction on an identified group manifold $\cX$ is reviewed. The coordinates of the higher-dimensional
space-time are $x^M=(x^{\mu}, y^i)$ where $y^i$ ($i=1,2,..d$) are coordinates on $\cX$ and $x^{\mu}$ ($\mu=d+1,d+2,..d+D$) are the coordinates
on the non-compact spacetime. The most general Einstein frame reduction ansatz invariant under rigid $G_L$ is
\begin{equation}\label{reduction ansatz}
d\widehat{s}^2=e^{2\alpha\varphi}ds^2_D+e^{2\beta\varphi}g_{mn}\nu^m\nu^n
\end{equation}
where the one-forms
\begin{equation}
\nu^m=\sigma^m - A^m
\end{equation}
include the Kaluza-Klein gauge fields $A^m_\m$, which have two-form field strength
\begin{equation}
F^m=dA^m+\frac{1}{2}f_{np}{}^mA^n\wedge A^p
\end{equation}
and $\a,\b$ in (\ref{reduction ansatz}) are the constants
\begin{eqnarray}
\alpha=-\left(\frac{d}{2(D-2)(D+d-2)}\right)^{\frac{1}{2}}  \qquad  \beta=-\frac{\alpha(D-2)}{d}
\end{eqnarray}
The left-invariant one-forms $\sigma^m=\sigma^m{}_i(y)dy^i$, where $m=1,2,..d$, satisfy the structure equation
\begin{equation}\label{structure eq}
d\sigma^m+\frac{1}{2}f_{np}{}^m\sigma^n\wedge\sigma^p=0
\end{equation}
which ensures that all $y^i$-dependence drops out of the reduced theory. The integrability condition $d^2\sigma^m=0$ gives the algebraic
constraint $f_{[mn}{}^qf_{p]q}{}^t=0$, and the invariance of the internal measure under $G_R$ requires that $G_R$ be
unimodular\footnote{Relaxing the unimodular condition still allows the reduction of the equations of motion, but not the Lagrangian.}
($f_{mn}{}^n=0$).

The $\nu^m$ define a covariant basis for the reduction in which the one forms transform under the local right action $G_R$, generated by the
globally defined left-invariant vector fields $Z_m=\sigma_m{}^i\partial_i$ as
\begin{equation}\label{graviphoton transformation}
\delta_Z(\omega)y^i=\omega^m\sigma_m{}^i    \qquad \delta_Z(\omega)\nu^m=-\nu^nf_{np}{}^m\omega^p  \qquad
\delta_Z(\omega)A^m=-d\omega^m-A^nf_{np}{}^m\omega^p
\end{equation}
Dimensional reduction gives rise to a metric $g_{\mu \nu}(x)$, $d$ Kaluza-Klein one-form gauge fields $A^m_\mu(x)$, and $d(d+1)/2$ scalars
$\varphi(x)$ and $g_{mn}(x)$, where $g_{mn}(x)$ is a positive definite symmetric matrix with unit determinant. This ansatz (\ref{reduction
ansatz}) is invariant under rigid $G_L$ transformations, and under local $G_R$ transformations in which the parameters depend on $x^\m$ and the
$A^m$ transform as gauge fields, while the scalar fields $g_{mn}(x)$ transform in the bi-adjoint.

The $D+d$-dimensional Einstein-Hilbert Lagrangian, reduced on a $d$-dimensional identified group manifold ${\cal X}_d$,  gives the effective
theory
\begin{eqnarray}\label{s-s lagrangian}
{\cal L}_D&=&R*1 - \frac{1}{2}*d\varphi\wedge d\varphi - \frac{1}{2}g^{mp}g^{nq}*Dg_{mn}\wedge Dg_{pq}
-\frac{1}{2}e^{2(\beta-\alpha)\varphi}g_{mn}*F^m \wedge F^n
\nonumber\\
&&-\frac{1}{4}e^{2(\alpha-\beta)\varphi}\left( g_{mn}g^{pq}g^{ts}f_{pt}{}^mf_{qs}{}^n+2g^{mn}f_{qm}{}^pf_{pn}{}^q\right)*1
\end{eqnarray}
where
\begin{eqnarray}
Dg_{mn}&=&dg_{mn}+g_{mp}f_{nq}{}^pA^q+g_{np}f_{mq}{}^pA^q
\end{eqnarray}
is a $G_R$-covariant derivative.

Theories of interest to us will also include antisymmetric tensor fields. In the ansatz of \cite{Scherk:1979zr}, the internal components
$T_{ij...k}$ of a tensor field $\widehat{T}_{MN...P}$ in the reduced theory are taken to have $y$ dependence given only by the frame fields
\begin{equation}
T_{ij...k}(x,y)=T_{mn...p}(x)\si _i{}^m\si _j{}^n...\si _k{}^p
\end{equation}
defining scalar fields $T_{mn...p}(x)$ in the reduced theory, so that for example the internal metric takes the form $g_{ij}(x,y)=g_{mn}(x)\si
_i{}^m\si _j{}^n$. As an example, consider the antisymmetric two-form potential, which we write in the ($G_R$-covariant) $\nu^m$ basis as
\begin{equation}
\widehat{B}_{(2)}=B_{(2)}+B_{(1)m}\wedge\nu^m+\frac{1}{2}B_{(0)mn}\nu^m\wedge\nu^n+\varpi_{(2)}
\end{equation}
where a left-invariant flux has been introduced
\begin{equation}
{\cal K}=\frac{1}{6}K_{mnp}\sigma^m\wedge\sigma^n\wedge\sigma^p
\end{equation}
where ${\cal K}=d\varpi_{(2)}$. The flux is closed (which requires the algebraic constraint $K_{[mn|p}f_{|qt]}{}^p=0$) but generally not exact
so that $\varpi_{(2)}$ is defined only locally. Defining the algebraic operator ${\cal Q}$, which acts on the space of antisymmetric tensors
$K_{m_1m_2...m_p}=K_{[m_1m_2...m_p]}$, such that
\begin{equation}
{\cal Q}:K_{m_1m_2...m_p}\rightarrow({\cal Q}K)_{m_1m_2...m_{p+1}}=f_{[m_1m_2}{}^nK_{m_3m_4...m_{p+1}]n}
\end{equation}
The condition $K_{[mn|p}f_{|qt]}{}^p=0$ may be written as $({\cal Q}K)_{mnpq}=0$, or $K_{mnp}\in Ker {\cal Q}$. The condition
$f_{[mn}{}^qf_{p]q}{}^t=0$ means that ${\cal Q}^2=0$ and we may define the algebraic cohomology $H({\cal Q})=Ker{\cal Q}/Im{\cal Q}$. It was
shown in \cite{Hull:2005hk} that a flux of the form $K_{mnp}=({\cal Q}\eta)_{mnp}$ for some $\eta_{mn}=-\eta_{nm}$ can be removed by a field
redefinition of $B$ and is therefore trivial. Therefore the fluxes of interest are those in $H({\cal Q})$. For field strengths with more
complicated Bianchi identities $dH\neq 0$, then $K_{[mn|p}f_{|qt]}{}^p\neq 0$ and the statement of algebraic cohomology must be suitably
modified.

The field strength $\widehat{H}_{(3)}=d\widehat{B}_{(2)}$ is invariant under the transformation
$\delta_X(\lambda)\widehat{B}_{(2)}=d\widehat{\lambda}_{(1)}$ where
\begin{equation}
\widehat{\lambda}_{(1)}=\lambda_{(1)}+\lambda_{(0)m}\nu^m
\end{equation}
and $\lambda_{(1)}=\lambda_{\mu}dx^{\mu}$. The antisymmetric tensor symmetry and the $G_R$ symmetry of $\cX$ gives the infinitesimal
transformations
\begin{eqnarray}
\delta B_{(2)}&=&d\lambda_{(1)}+\frac{1}{2}K_{mnp}\omega^pA^m\wedge A^n\nonumber\\
\delta
B_{(1)m}&=&D\lambda_{(0)m}+B_{(1)n}f_{mp}{}^n\omega^p-K_{mnp}\omega^pA^n\nonumber\\
\delta B_{(0)mn}&=&f_{mn}{}^p\lambda_{(0)p}+2B_{(0)[m|p}f_{|n]q}{}^p\omega^q+K_{mnp}\omega^p
\end{eqnarray}
where $D\lambda_{(0)m}=d\lambda_{(0)m}+f_{mn}{}^p\lambda_{(0)p}A^n$. Combining these variations with that of the graviphoton $\delta
A^m=-D\omega^m$, these infinitesimals generate the Lie algebroid
\begin{eqnarray}
\label{B-field algebra} \lbrack \delta_{Z}(\widetilde{\omega}),\delta_{Z}(\omega) \rbrack&=&\delta_{Z}(f_{np}{}^m\omega^n\widetilde{\omega}^p) -
\delta_{X}(K_{mnp}\omega^n\widetilde{\omega}^p) - \delta_{W}(K_{mnp}\omega^n\widetilde{\omega}^pA^m)
\nonumber\\
\lbrack \delta_{X}(\lambda),\delta_{Z}(\omega) \rbrack&=&-\delta_{X}(\lambda_mf_{np}{}^m\omega^p)\nonumber\\
\lbrack \delta_{X}(\widetilde{\lambda}),\delta_{X}(\lambda) \rbrack&=&0
\end{eqnarray}
 where $\delta_{Z}(\omega)=\omega^mZ_m$, $\delta_{X}(\lambda)=\lambda_{(0)m}X^m$ and
$\delta_{W}(\lambda)=\lambda_{\mu}W^{\mu}$. As argued in \cite{Hull:2005hk,Hull:2006tp}, such field dependence is characteristic of theories in
which we require field strengths to have Chern-Simons terms in order to be gauge invariant and such Chern-Simons terms are generated naturally
by dimensional reduction. The algebra may be viewed as a Lie algebra bundle over the non-compact $D$-dimensional spacetime, i.e. at each point
$x_o$ on the base, the graviphoton is constant along the fibre and the algebroid reduces to a Lie algebra with structure constants $f_{mn}{}^p$,
$K_{mnp}$ and $K_{mnp}A^p_{\mu}(x_o)$.

The algebroid (\ref{B-field algebra}) has Lie subalgebra \cite{Odd}
\begin{eqnarray}
\left[Z_m,Z_n\right]&=&-f_{mn}{}^pZ_p+K_{mnp}X^p\nonumber\\
\left[X^m,Z_n\right]&=&f_{np}{}^mX^p\nonumber\\
\left[X^m,X^n\right]&=&0
\end{eqnarray}
In the next section we shall consider a generalisation of this reduction to the full bosonic sector of the IIB supergravity.

\section{Compactifications of IIB Supergravity}

We consider here the reduction of the bosonic sector of IIB supergravity on identified group manifolds with flux. The general reduction of the
Fermi sector and Supersymmetry breaking will be considered elsewhere\footnote{See \cite{Grana:2006kf} for a recent discussion of $\mathcal{N}=1$
vacua in the case where $\mathcal{X}$ is a six-dimensional Nil or Solve-manifold.}. An important property of the IIB theory is the self-duality
of the Ramond-Ramond five-form field strength.
\begin{equation}\label{self duality constraint}
\widehat{G}_{(5)}=*\widehat{G}_{(5)}
\end{equation}
Such a constraint cannot naturally be encoded in a Lagrangian formalism and it must be separately imposed on the equations of motion. The
approach will be to treat $\widehat{G}_{(5)}$ and $*\widehat{G}_{(5)}$ as independent fields in the Lagrangian and impose the self-duality
constraint after the dimensional reduction. We shall only be interested in the general structure of the reduced theory, in particular the gauge
symmetries, so the issue of self duality will not play a significant role.

The bosonic sector of the ten-dimensional IIB Lagrangian, written in a manifestly $SL(2)$ invariant form is
\begin{equation}\label{IIB Lagrangian}
{\cal L}_{IIB}=\widehat{{\cal R}}*1+\frac{1}{4}tr\left(*d\widehat{{\cal K}} \wedge d\widehat{{\cal K}}^{-1}\right)-\frac{1}{2}\widehat{{\cal
K}}_{a b}*\widehat{\cal H}^{a}_{(3)} \wedge \widehat{\cal H}^{b}_{(3)}-\frac{1}{4}*\widehat{G}_{(5)} \wedge
\widehat{G}_{(5)}-\frac{1}{4}\epsilon_{a b}\widehat{\cal C}_{(4)} \wedge \widehat{\cal H}_{(3)}^{a} \wedge \widehat{\cal H}^{b}_{(3)}
\end{equation}
Where the 3-form field strengths ${\cal H}^a_{(3)}$ transform as a doublet under $SL(2)$, the self-dual five-form $\widehat{G}_{(5)}$ as a
singlet and the axio-dilaton $\tau$ (written above in terms of the scalars $\widehat{\cal K}$) in a fractionally linear way
\begin{equation}
\tau\rightarrow\frac{a\tau+b}{c\tau+d}  \qquad  \left(%
\begin{array}{cc}
  a & b \\
  c & d \\
\end{array}%
\right)\in SL(2)
\end{equation}
The trace in (\ref{IIB Lagrangian}) is taken over the $SL(2)$ indices $a=1,2$. The field strengths and scalars $\widehat{\cal K}$ are defined;
\begin{equation}\label{definitions}
\widehat{G}_{(5)}=d\widehat{\cal C}_{(4)}+\frac{1}{2}\epsilon_{a b}\widehat{\cal B}^{a}_{(2)}\wedge
\widehat{\cal H}^{b}_{(3)} \qquad \widehat{\cal H}^{a}_{(3)}=d\widehat{\cal B}_{(2)}^{a}=\left(\begin{array}{c} db_{(2)} \\
dc_{(2)}
\end{array}\right)\qquad \widehat{{\cal K}}=e^{\phi} \left(\begin{array}{cc}
1 & C_{(0)} \\ C_{(0)} & |\tau|^2
\end{array}\right)
\end{equation}
where
\begin{equation} \epsilon_{ab}= \left(\begin{array}{cc}
0 & 1 \\ -1 & 0
\end{array}\right) \qquad \tau=C_{(0)}+ie^{-\phi}
\end{equation}
The scalar sector consists of a dilaton $\phi$ and a Ramond-Ramond zero-form $C_{(0)}$ which parameterise the coset manifold
$SL(2;\R)/SO(2)\simeq SU(1,1)/U(1)$. $\widehat{\cal C}_{(4)}$ and $c_{(2)}$ are p-form fields arising from the massless Ramond-Ramond sector of
the Type IIB String spectrum and $b_{(2)}$ is the Kalb-Ramond potential. The ten dimensional Equations of motion derived from the Lagrangian
(\ref{IIB Lagrangian}) are
\begin{eqnarray}\label{equations of motion}
d*\widehat{G}_{(5)}&=&\frac{1}{2}\epsilon_{ab}\widehat{H}^a_{(3)}\wedge\widehat{H}^b_{(3)}\nonumber\\
d*\widehat{{\cal K}}_{ab}\widehat{H}^a_{(3)}&=&\epsilon_{ab}\widehat{H}^b_{(3)}\wedge\widehat{G}_{(5)}\nonumber\\
d*d\widehat{{\cal K}}^{ab}&=&*\widehat{H}^a_{(3)}\wedge\widehat{H}^b_{(3)}
\end{eqnarray}
and the Bianchi identities are
\begin{equation}\label{bianchi identities}
d\widehat{G}_{(5)}-\frac{1}{2}\epsilon_{ab}\widehat{{\cal H}}^a\wedge\widehat{{\cal H}}^b=0   \qquad  d\widehat{{\cal H}}^a=0
\end{equation}
which are consistent with the self-duality constraint (\ref{self duality constraint}). The action of S-duality on these fields is
\begin{equation}\label{s duality}
\widehat{{\cal K}}\rightarrow S^t\widehat{{\cal K}}S    \qquad \widehat{\cal B}_{(2)}\rightarrow S^{-1}\widehat{\cal B}_{(2)}
\end{equation}
and $\epsilon_{ab}$ is invariant
\begin{equation}
\epsilon\rightarrow S^t\epsilon S=\epsilon
\end{equation}
where $S\in SL(2;\Z)\subset SL(2)$.

\subsection{Inclusion of Fluxes}

The flux ansatz for the two form is a generalisation of that described in \cite{Hull:2005hk,Odd}, which transforms covariantly under S-duality,
mixing the Neveu-Schwarz and Ramond two form fluxes. A left-invariant flux, $M_{(3)}^a$, is included in the two form reduction via the ansatz
\begin{equation}
\widehat{\cal B}^a_{(2)}=\widehat{B}^a_{(2)} + \varpi_{(2)}^a \qquad \widehat{\cal H}^a_{(3)}=\widehat{H}^a_{(3)} + M_{(3)}^a
\end{equation}
where $\widehat{H}^a_{(3)}=d\widehat{B}^a_{(2)}$ and
\begin{equation}
d\varpi_{(2)}^a=M^a_{(3)}= \frac{1}{6}M_{mnp}{}^a \sigma^m \wedge \sigma^n \wedge \sigma^p
\end{equation}
$M_{mnp}{}^a$ are constant, $SL(2)$ valued antisymmetric coefficients. Introducing flux on the $\widehat{G}_{(5)}$ field strength is not
straightforward due to the Chern-Simons term in $\widehat{G}_{(5)}$. Such terms threaten the consistency of the truncation as they introduce
bare flux potential terms $\varpi_{(2)}^a$ which have explicit $y$ dependence. In \cite{Odd} it was demonstrated that (for the Heterotic
string), by a careful choice of flux, the consistency of the truncation may be maintained even for theories with such Chern-Simons terms. These
techniques can be generalised to higher degree forms and applied to the $\widehat{\cal C}_{(4)}$ potential of the IIB theory to give the flux
ansatz
\begin{equation}\label{four form flux def}
\widehat{\cal C}_{(4)} = \widehat{S}_{(4)} -\frac{1}{2}\epsilon_{ab}\varpi_{(2)}^a \wedge \widehat{B}_{(2)}^b +\varpi_{(4)}
\end{equation}
where
\begin{equation}\label{4 Flux}
d\varpi_{(4)}=-\frac{1}{2}\epsilon_{ab}\varpi_{(2)}^a\wedge M_{(3)}^b+{\cal K}_{(5)}
\end{equation}
${\cal K}_{(5)}$ is the left-invariant five-form flux
\begin{equation}
{\cal K}_{(5)}=\frac{1}{120}K_{mnpqt} \sigma^m \wedge \sigma^n \wedge \sigma^p \wedge \sigma^q \wedge \sigma^t
\end{equation}
The first term on the right hand side in (\ref{4 Flux}) is required to cancel any $y$-dependance in the five form field strength that may arise
due to the flux on $\widehat{\cal B}^a_{(2)}$ in the Chern-Simons term of $\widehat{G}_{(5)}$. The reduction ansatz for the five-form field
strength (\ref{definitions}) is
\begin{eqnarray}
\widehat{G}_{(5)}&=&d\widehat{S}_{(4)}+\frac{1}{2}\epsilon_{ab}\left(\widehat{B}^a_{(2)} \wedge \widehat{H}^b_{(3)} - 2 M_{(3)}^a \wedge
\widehat{B}^b_{(2)}\right) + {\cal K}_{(5)}
\end{eqnarray}

It will be important, especially when considering the symmetry transformations, to distinguish between those parts of the potentials with flux
included implicitly, denoted by the calligraphic script $\widehat{\cal B}^a_{(2)}$ and $\widehat{\cal C}_{(4)}$, and those without,
$\widehat{B}^a_{(2)}$ and $\widehat{S}_{(4)}$. The requirement that the fluxes do not alter the Bianchi identities (\ref{bianchi identities})
requires that the fluxes satisfy
\begin{equation}
d\left(\frac{1}{6}M_{mnp}{}^a\sigma^m\wedge\sigma^n\wedge\sigma^p\right)=0  \qquad d\left(-\frac{1}{2}\epsilon_{ab}\varpi_{(2)}^a\wedge
M_{(3)}^b+{\cal K}_{(5)}\right)=0
\end{equation}
which impose the algebraic conditions
\begin{eqnarray}\label{flux constraints}
M_{[mn|t}{}^af_{|pq]}{}^t&=&0\nonumber\\
2\epsilon_{ab}M_{[mnp}{}^aM_{qts]}{}^b+3K_{[mnpq|l}f_{|ts]}{}^l&=&0
\end{eqnarray}
In addition to the condition $f_{[mn}{}^qf_{p]q}{}^t=0$.

\subsection{Flux Compactification on Identified Group Manifolds}

The Chern-Simons term of the ten-dimensional IIB Lagrangian has an explicit dependence on the potential of the fluxes $\varpi^a_{(2)}$ and
$\varpi_{(4)}$, entering through $\widehat{\cal C}_{(4)}$. It is the fluxes $M^a_{(3)}$ and ${\cal K}_{(5)}$ that are globally defined, not the
potentials so one might worry that the Lagrangian is not well defined and the reduction not consistent. However variation of the Lagrangian with
respect to the potentials $\widehat{S}_{(4)}$ and $\widehat{B}^a_{(2)}$ still yield the correct, well defined, equations of motion
(\ref{equations of motion}). One way to proceed would be to disregard the Lagrangian (\ref{IIB Lagrangian}) and reduce the equations of motion
(\ref{equations of motion}) directly.

The fact that the physics depends only on the fluxes $M^a_{(3)}$ and ${\cal K}_{(5)}$ and not the potentials $\varpi^a_{(2)}$ and $\varpi_{(4)}$
is due to the gauge invariance of the theory under antisymmetric tensor transformations. However, even though the equations of motion are
manifestly invariant under the tensor transformations, the Lagrangian is not. Consider the Chern-Simons form contribution to the action
\begin{equation}
S_{CS}=\frac{1}{4}\int_M\epsilon_{ab}\widehat{\cal C}_{(4)} \wedge \widehat{\cal H}_{(3)}^{a} \wedge \widehat{\cal H}^{b}_{(3)}
\end{equation}
under the antisymmetric tensor transformation $\delta_X \widehat{\cal C}=d\widehat{\Lambda}$ the action transforms as
\begin{eqnarray}
\delta_X S_{CS}&=&\frac{1}{4}\int_M\epsilon_{ab}d\widehat{\Lambda} \wedge \widehat{\cal H}_{(3)}^{a} \wedge \widehat{\cal H}^{b}_{(3)}\nonumber\\
&=&\frac{1}{4}\int_{\partial M}\epsilon_{ab}\widehat{\Lambda} \wedge \widehat{\cal H}_{(3)}^{a} \wedge \widehat{\cal H}^{b}_{(3)}
\end{eqnarray}
which is zero if either the ten-dimensional spacetime has no boundary or $\widehat{\Lambda}$ vanishes on the boundary. The problem arises when
one considers large gauge transformations. The fact that $S_{CS}$ is not manifestly invariant under large gauge transformations is related to
the appearance of the bare flux potentials $\varpi_{(2)}^a$ and $\varpi_{(4)}$. This issue, in addition to the self-duality constraint provides
good motivation to consider the reductions of the equations of motion directly. However there are some issues, such as moduli fixing, in which
it can be helpful to have an explicit reduction of the scalar potential.

The left-invariant Scherk-Schwarz reduction ansatze are
\begin{eqnarray}\label{C def}
\widehat{S}_{(4)}&=&S_{(4)}+S_{(3)m}\wedge \nu^m+\frac{1}{2}S_{(2)mn}\wedge \nu^m\wedge \nu^n+\frac{1}{6}S_{(1)mnp}\wedge \nu^m\wedge
\nu^n\wedge \nu^p\nonumber\\&&+\frac{1}{24}S_{(0)mnpq}\nu^m\wedge \nu^n\wedge \nu^p\wedge \nu^q
\nonumber\\
\widehat{G}_{(5)}&=&G_{(5)}+G_{(4)m}\wedge \nu^m+\frac{1}{2}G_{(3)mn}\wedge \nu^m\wedge \nu^n+\frac{1}{6}G_{(2)mnp}\wedge \nu^m\wedge
\nu^n\wedge \nu^p\nonumber\\&&+\frac{1}{24}G_{(1)mnpq}\wedge\nu^m\wedge \nu^n\wedge \nu^p\wedge \nu^q+\frac{1}{120}G_{(0)mnpqt}\nu^m\wedge
\nu^n\wedge \nu^p\wedge \nu^q\wedge\nu^t\nonumber\\
\end{eqnarray}
where $\widehat{G}_{(5)}$ is defined by (\ref{definitions}) For the two-form we define
\begin{eqnarray}\label{B def}
\widehat{B}^a_{(2)}&=&B^a_{(2)}+B^a_{(1)m}\wedge \nu^m+\frac{1}{2}B^a_{(0)mn}\nu^m\wedge \nu^n
\nonumber\\
\widehat{H}^a_{(3)}&=&H^a_{(3)}+H^a_{(2)m}\wedge \nu^m+\frac{1}{2}H^a_{(1)mn}\wedge\nu^m\wedge
\nu^n+\frac{1}{6}H^a_{(0)mnp}\nu^m\wedge\nu^n\wedge\nu^p\nonumber\\
\end{eqnarray}
and similarly for the three form field strength with flux $\widehat{{\cal H}}^a_{(3)}$
\begin{equation}
\widehat{{\cal H}}^a_{(3)}={\cal H}^a_{(3)}+{\cal H}^a_{(2)m}\wedge\nu^m+\frac{1}{2}{\cal H}^a_{(1)mn}\wedge\nu^m\wedge\nu^n+\frac{1}{6}{\cal
H}^a_{(0)mnp}\nu^m\wedge\nu^n\wedge\nu^p
\end{equation}
The reduced field strengths and Bianchi identities are given in Appendix A. The reduced theory has scalar potential
 \begin{eqnarray}
V&=&-\frac{1}{2}e^{2(\beta-\alpha)\varphi}\left(
g_{mn}g^{pq}g^{ts}f_{pt}{}^mf_{qs}{}^n+2g^{mn}f_{qm}{}^pf_{pn}{}^q\right)\nonumber\\
&&-\frac{1}{4}e^{-10(\beta-\alpha)\varphi}g^{mn}g^{pq}g^{ts}g^{lk}g^{ij}G_{(0)mptli} G_{(0)nqskj}\nonumber\\&&-
\frac{1}{2}e^{-6(\beta-\alpha)\varphi}g^{mn}g^{pq}g^{ts}{\cal K}_{ab}\mathcal{H}^a_{(0)mpt}\mathcal{H}^b_{(0)nqs}
 \end{eqnarray}

\subsection{Gauge Symmetry}

In this section the gauge symmetries of the IIB theory reduced on an identified group manifold ${\cal X}$ with flux described in the previous
sections are investigated. The presence of the Chern-Simons term in $\widehat{G}_{(5)}$ leads to a gauge algebra with a far more complicated
structure than seen in (\ref{B-field algebra}).

\subsubsection{Three Form Anti-Symmetric Tensor Transformations}

The antisymmetric tensor transformation $\widehat{S}_{(4)}\rightarrow \widehat{S}_{(4)}+d\widehat{\Lambda}_{(3)}$ leaves the field strength
$\widehat{G}_{(5)}$ invariant and is generated by the parameters
\begin{equation}
\widehat{\Lambda}_{(3)}=\Lambda_{(3)}+\Lambda_{(2)m}\wedge\nu^m
+\frac{1}{2}\Lambda_{(1)mn}\wedge\nu^m\wedge\nu^n+\frac{1}{6}\Lambda_{(0)mnp}\nu^m\wedge\nu^n\wedge\nu^p
\end{equation}
Gauge transformations with respect to each component of $\widehat{\Lambda}_{(3)}$ are generated by $\delta_X(\Lambda_{(3)})$,
$\delta_X(\Lambda_{(2)m})$, $\delta_X(\Lambda_{(1)mn})$ and $\delta_X(\Lambda_{(0)mnp})$. The generators of this transformation are defined as
$\delta_X(\widehat{\Lambda})$ and their action on the reduced potential is
\begin{eqnarray}\label{X-transformations}
\delta_X(\widehat{\Lambda})S_{(4)}&=&d\Lambda_{(3)}-\Lambda_{(2)m}\wedge F^m
\nonumber\\
\delta_X(\widehat{\Lambda})S_{(3)m}&=&D\Lambda_{(2)m}+\Lambda_{(1)mn}\wedge F^n
\nonumber\\
\delta_X(\widehat{\Lambda})S_{(2)mn}&=&-\Lambda_{(2)p}f_{mn}{}^p+D\Lambda_{(1)mn}-\Lambda_{(0)mnp}F^p
\nonumber\\
\delta_X(\widehat{\Lambda})S_{(1)mnp}&=&O_{mnp}^{qt}\Lambda_{(1)qt}+D\Lambda_{(0)mnp}
\nonumber\\
\delta_X(\widehat{\Lambda})S_{(0)mnpq}&=&-O_{mnpq}^{tsl}\Lambda_{(0)tsl}
\end{eqnarray}
where constants $O_{mnp}^{qt}$ and $O_{mnpq}^{tsl}$ are defined as
\begin{eqnarray}
O_{mnp}^{qt}&=&3\delta^q_{[m}f_{np]}{}^t\nonumber\\
O_{mnpq}^{tsl}&=&6\delta_{[m}{}^t\delta_{n}{}^sf_{pq]}{}^l
\end{eqnarray}

\subsubsection{One-Form Anti-symmetric Tensor Transformations}

Consider the symmetry generated by the gauge transformation
\begin{equation}
\delta_Y(\widehat{\lambda})\widehat{\cal B}^a_{(2)}=\delta_Y(\widehat{\lambda})\widehat{B}^a_{(2)}=d\widehat{\lambda}^a_{(1)}
\end{equation}
The three form field strength $\widehat{\cal H}^a_{(3)}$ is manifestly invariant under this transformation, but invariance of the five form
$\widehat{G}_{(5)}$ requires a compensating transformation from $\widehat{S}_{(4)}$. The self-dual five form field strength is
\begin{equation}
\widehat{G}_{(5)} = d\widehat{S}_{(4)}+\frac{1}{2} \epsilon_{ab} \left(\widehat{B}^a_{(2)} \wedge d\widehat{B}^b_{(2)}-2M_{(3)}^a \wedge
\widehat{B}^b_{(2)} \right) + {\cal K}_{(5)}
\end{equation}
We define the effect of the infinitesimal transformation $\delta_Y(\widehat{\lambda})$ on $\widehat{S}_{(4)}$ as that which ensures;
$\delta_Y(\widehat{\lambda}) \widehat{G}_{(5)}=0$, i.e.
\begin{equation}
\delta_Y(\widehat{\lambda})\widehat{G}_{(5)} = d\left(\delta_Y(\widehat{\lambda})\widehat{S}\right)+\frac{1}{2} \epsilon_{ab}\left(
d\widehat{\lambda}^a_{(1)} \wedge d\widehat{B}^b_{(2)}- 2M_{(3)}^a \wedge d\widehat{\lambda}^b_{(1)}\right) =0
\end{equation}
Integrating this equation gives an expression for the gauge transformation of $\widehat{S}_{(4)}$;
\begin{equation}\label{one form gauge transformation}
\delta_Y(\widehat{\lambda})\widehat{S}_{(4)}=-\frac{1}{2} \epsilon_{ab}\left(d\widehat{\lambda}^a_{(1)}\wedge \widehat{B}^b_{(2)}+2
\widehat{\lambda}^a_{(1)}\wedge M_{(3)}^b \right) + d\Lambda_{(3)}
\end{equation}
where $\Lambda_{(3)}$ is an arbitrary 3-form, which we shall set to zero. The gauge parameter and its exterior derivative are
\begin{eqnarray}\label{gauge parameter}
\widehat{\lambda}_{(1)}^a&=&\lambda^a_{(1)}+\lambda^a_{(0)m}\nu^m
\nonumber\\
d\widehat{\lambda}^a_{(1)}&=&d\lambda^a_{(1)}-\lambda^a_{(0)m} F^m + D\lambda_{(0)m}^a\wedge \nu^m - \frac{1}{2}\lambda^a_{(0)p}f_{mn}{}^p\nu^m
\wedge\nu^n
\end{eqnarray}
Substituting (\ref{gauge parameter}) in (\ref{one form gauge transformation}), the one-form gauge transformations are
\begin{eqnarray}
\delta_Y(\widehat{\lambda})S_{(4)}&=&-\frac{1}{2}\epsilon_{ab}\left(d\lambda^a_{(1)}-\lambda^a_{(0)m}F^m\right)\wedge
B^b_{(2)}-\frac{1}{6}\epsilon_{ab}\lambda^a_{(1)} M_{mnp}{}^b\wedge A^m\wedge A^n\wedge A^p
\nonumber\\\nonumber\\
\delta_Y(\widehat{\lambda})S_{(3)m}&=&-\frac{1}{2}\epsilon_{ab}\left(d\lambda^a_{(1)}-\lambda^a_{(0)n}F^n\right)\wedge
B^b_{(1)m}-\frac{1}{2}\epsilon_{ab}\lambda^a_{(1)} M_{mnp}{}^b\wedge A^n\wedge A^p\nonumber\\&& -
\frac{1}{2}\epsilon_{ab}D\lambda^a_{(0)m}\wedge B^b_{(2)} - \frac{1}{6}\epsilon_{ab}\lambda^a_{(0)m}M_{npq}{}^bA^n\wedge A^p\wedge A^q
\nonumber\\\nonumber\\
\delta_Y(\widehat{\lambda})S_{(2)mn}&=&-\frac{1}{2}\epsilon_{ab}\left(d\lambda^a_{(1)}-\lambda^a_{(0)p}F^p\right)
B^b_{(0)mn}-\epsilon_{ab}\lambda^a_{(1)}\wedge M_{mnp}{}^bA^p\nonumber\\&&+\frac{1}{2}\epsilon_{ab}\lambda^a_{(0)p}f_{mn}{}^pB^b_{(2)}
-\epsilon_{ab}D\lambda^a_{(0)[m}\wedge B^b_{(1)n]} -\epsilon_{ab}\lambda^a_{(0)[m}M_{n]pq}{}^bA^p\wedge A^q
\nonumber\\\nonumber\\
\delta_Y(\widehat{\lambda})S_{(1)mnp}&=&-\epsilon_{ab}\lambda^a_{(1)}M_{mnp}{}^b-\frac{3}{2}\epsilon_{ab}D\lambda^a_{(0)[m}B^b_{(0)np]}
-3\epsilon_{ab}\lambda^a_{(0)[m}M_{np]q}{}^bA^q\nonumber\\
&&+\frac{3}{2}\epsilon_{ab}\lambda_{(0)q}^af_{[mn}{}^qB^b_{(1)|p]}
\nonumber\\\nonumber\\
\delta_Y(\widehat{\lambda})S_{(0)mnpq} &=& 6\epsilon_{ab}\lambda^a_{(0)t}f_{[mn}{}^tB^b_{(0)pq]} - 4\epsilon_{ab} \lambda^a_{(0)[m}M_{npq]}{}^b
\end{eqnarray}
and for the two form
\begin{eqnarray}
\delta_Y(\widehat{\lambda})B^a_{(2)}&=&d\lambda^a_{(1)}-\lambda^a_{(0)m}F^m
\nonumber\\
\delta_Y(\widehat{\lambda})B^a_{(1)m}&=&D\lambda^a_{(0)m}
\nonumber\\
\delta_Y(\widehat{\lambda})B^a_{(0)mn}&=&-\lambda^a_{(0)p}f_{mn}{}^p
\end{eqnarray}

\subsubsection{Right Action of the Group Manifold}

$\cX=G/\G$ inherits the right action of the group $G_R$ on $G$. The calculation of how the reduced fields transform under $G_R$ is somewhat
involved and only the results are given here. Details of the calculation may be found in Appendix B. The right action gives
\begin{eqnarray}\label{C diffeomorphism 2}
\delta_Z(\omega^m)S_{(4)}&=&\frac{1}{4}\epsilon_{ab}M_{mnp}{}^a\omega^pB_{(2)}^b\wedge A^m\wedge A^n-\frac{1}{24}K_{mnpqt}\omega^tA^m\wedge
A^n\wedge A^p\wedge A^q
\nonumber\\\nonumber\\
\delta_Z(\omega^m)S_{(3)m}&=&S_{(3)n}f_{mp}{}^n\omega^p+\frac{1}{4}\epsilon_{ab}M_{npq}{}^a\omega^qA^n\wedge A^p\wedge
B^b_{(1)m}\nonumber\\
&&+\frac{1}{2}\epsilon_{ab}M_{mnp}{}^a\omega^pA^n\wedge B^b_{(2)} +\frac{1}{6}K_{mnpqt}\omega^tA^n\wedge A^p\wedge A^q
\nonumber\\\nonumber\\
\delta_Z(\omega^m)S_{(2)mn}&=&2S_{(2)[m|p}f_{|n]q}{}^p\omega^q+\frac{1}{4}\epsilon_{ab}M_{pqt}{}^a\omega^tA^p\wedge
A^qB_{(0)mn}^b+\epsilon_{ab}M_{mpq}{}^a\omega^pA^q\wedge
B^b_{(1)n}\nonumber\\
&&+\frac{1}{2}\epsilon_{ab}M_{mnp}{}^a\omega^pB^b_{(2)} -\frac{1}{2}K_{mnpqt}\omega^tA^p\wedge A^q
\nonumber\\\nonumber\\
\delta_Z(\omega^m)S_{(1)mnp}&=&3S_{(1)[mn|q}f_{|p]t}{}^q\omega^t+ \frac{3}{2}\epsilon_{ab}M_{mnq}{}^a\omega^q B^b_{(1)p}
-\frac{3}{2}\epsilon_{ab}M_{mqt}{}^a\omega^t B^b_{(0)np}A^q\nonumber\\
&&+K_{mnpqt}\omega^tA^q
\nonumber\\\nonumber\\
\delta_Z(\omega^m)S_{(0)mnpq}&=&4S_{(0)[mnp|t}f_{|q]s}{}^t\omega^s+3\epsilon_{ab}M_{[mn|t}{}^aB^b_{(0)|pq]}\omega^t-K_{mnpqt}\omega^t
\end{eqnarray}

\subsubsection{Gauge Algebra}

Using the results of the previous sections, the full gauge algebra of the compactified IIB theory is
\begin{eqnarray}\label{IIB algebra}
\left[\delta_Z\left(\bar{\omega}\right),\delta_Z\left(\omega\right)\right]&=&
\delta_Z\left(f_{np}{}^m\omega^n\bar{\omega}^p\right)-\delta_X\left(K_{mnpqt}\omega^q\bar{\omega}^t\right)
-\delta_X\left(K_{mnpqt}\omega^q\bar{\omega}^tA^p\right) \nonumber\\&&-\delta_X\left(\frac{1}{2}K_{mnpqt}\omega^q\bar{\omega}^tA^n\wedge
A^p\right)\nonumber\\&&-\delta_X\left(\frac{1}{6}K_{mnpqt}\omega^q\bar{\omega}^tA^m\wedge A^n\wedge A^p\right)
\nonumber\\&&-\delta_Y\left(M_{mnp}{}^a\omega^n\bar{\omega}^p\right) -\delta_Y\left(M_{mnp}{}^a\omega^n\bar{\omega}^pA^m\right)
\nonumber\\\nonumber\\
\left[\delta_X(\Lambda_{(2)m}),\delta_Z(\omega^m)\right]&=&\delta_X\left(\Lambda_{(2)n}f_{mp}{}^n\omega^p\right)
\nonumber\\\nonumber\\
\left[\delta_X(\Lambda_{(1)mn}),\delta_Z(\omega^q)\right]&=&\delta_X\left(\Lambda_{(1)mp}f_{nq}{}^p\omega^q\right)
-\delta_X\left(\Lambda_{(1)np}f_{mq}{}^p\omega^q\right)
\nonumber\\\nonumber\\
\left[\delta_X(\Lambda_{(0)mnp}),\delta_Z(\omega^m)\right]&=&\delta_X\left(\Lambda_{(0)mnq}f_{pt}{}^q\omega^t\right)
+\delta_X\left(\Lambda_{(0)npq}f_{mt}{}^q\omega^t\right)+\delta_X\left(\Lambda_{(0)pmq}f_{nt}{}^q\omega^t\right)
\nonumber\\\nonumber\\
\left[\delta_Y(\lambda_{(1)}^a),\delta_Z(\omega^m)\right]&=& -\delta_X\left(\epsilon_{ab}M_{mnp}{}^b\omega^p\lambda_{(1)}^b\right)
-\delta_X\left(\epsilon_{ab}M_{mnp}{}^a\omega^p\lambda^b_{(1)}\wedge A^n\right)
\nonumber\\&&-\delta_X\left(\frac{1}{2}\epsilon_{ab}M_{mnp}{}^a\omega^p\lambda^b_{(1)}\wedge A^m\wedge A^n\right)
\nonumber\\\nonumber\\
\left[\delta_Y(\lambda_{(0)m}^a),\delta_Z(\omega^m)\right]&=&
\delta_Y\left(\lambda^a_{(0)n}f_{mp}{}^n\omega^p\right)-\delta_X\left(3\epsilon_{ab}\lambda^a_{(0)m}M_{npq}{}^b\omega^q\right)\nonumber\\
\end{eqnarray}
All other commutators vanish. This gauge algebra contains a Lie algebra subgroup. A naive guess for the Lie algebra is
\begin{eqnarray}\label{short lie algebra}
\left[Z_m,Z_n\right]&=&-f_{mn}{}^pZ_p-M_{mnp}{}^aY_a{}^p-K_{mnpqt}X^{pqt}
\nonumber\\
\left[X^{mnp},Z_q\right]&=&3f_{qt}{}^{[m}X^{np]t}
\nonumber\\
\left[Y_a{}^m,Z_n\right]&=&f_{np}{}^mY_a{}^p-3\epsilon_{ab}M_{npq}{}^bX^{mpq}
\end{eqnarray}
with all other commutators vanishing. As in the eleven dimensional supergravity case \cite{Hull:2006tp}, the the Jacobi identity for this
algebra fails to close and a truncation of the set of generators must be considered. This is a consequence of the reducibility of the gauge
transformations. It will be shown in the examples of section 3.4 that the irreducible gauge transformations correspond to the irreducible
representations of the gauge group. Consider for example the slightly simpler case where $M_{mnp}{}^a=0$ then algebra (\ref{short lie algebra})
reduces to
\begin{eqnarray}\label{very short algebra}
\left[Z_m,Z_n\right]&=&-f_{mn}{}^pZ_p-K_{mnpqt}X^{pqt}
\nonumber\\
\left[X^{mnp},Z_q\right]&=&f_{qt}{}^mX^{npt}+f_{qt}{}^pX^{mnt}+f_{qt}{}^nX^{pmt}
\end{eqnarray}
The triple commutator for $Z_m$ is
\begin{equation}
[[Z_m,Z_n],Z_p]+[[Z_n,Z_p],Z_m]+[[Z_p,Z_m],Z_n]=3K_{mnp[q|t}f_{|sl]}{}^tX^{qsl}=K_{mnpjt}O_{qsl}^{jt}X^{qsl}\neq 0
\end{equation}
where the constants $O_{mnp}^qt$ and $\Pi^{mnp}_{qt}$ are
\begin{eqnarray}
O^{qt}_{mnp}&=&3\delta^q{}_{[m}f_{np]}{}^t \nonumber\\
\Pi^{mnp}_{qt}&=&\frac{1}{2}\delta^{[m}{}_qf_t{}^{np]}
\end{eqnarray}
We see that the commutators (\ref{very short algebra}) do not satisfy the Jacobi identity and is therefore not a Lie algebra. The apparent
non-associativity of (\ref{very short algebra}) may be understood by considering the following example. For simplicity consider the case where
the group $G$ the identified group manifold is constructed from is chosen to be semi-simple. It is useful to decompose the generator $X^{mnp}$
as\footnote{Consider the further decomposition, $\widetilde{X}^{qt}\rightarrow\widetilde{X}^{qt}+f_s{}^{qt}\widetilde{X}^s$, where
$f_{mn}{}^p\widetilde{X}^{mn}=0$. The action of $\Pi^{mnp}_{qt}$ projects out the $\widetilde{X}^s$ contribution, since
\begin{equation}
\Pi^{mnp}_{qt}\widetilde{X}^{qt}\rightarrow
\Pi^{mnp}_{qt}\widetilde{X}^{qt}+\frac{1}{2}f_t^{[np}f_s{}^{m]t}\widetilde{X}^s=\Pi^{mnp}_{qt}\widetilde{X}^{qt}
\end{equation}
Therefore, for our purposes, this second decomposition need not be explicitly stated, except to note that $f_{mn}{} ^p\widetilde{X}^{mn}=0$.}
\begin{equation}\label{decomposition}
X^{mnp}=\widetilde{X}^{mnp}+\Pi^{mnp}_{qt}\widetilde{X}^{qt}
\end{equation}
where $O_{mnp}^{qt}\widetilde{X}^{mnp}=0$ and $f_{np}{}^m\widetilde{X}^{np}=0$, such that $O_{mnp}^{qt}X^{mnp}=\widetilde{X}^{qt}$. The algebra
\begin{eqnarray}
\left[Z_m,Z_n\right]&=&-f_{mn}{}^pZ_p-K_{mnpqt}\widetilde{X}^{pqt}
\nonumber\\
\left[\widetilde{X}^{mnp},Z_q\right]&=&3f_{qt}{}^{[m}\widetilde{X}^{np]t}
\end{eqnarray}
satisfies the Jacobi identity and \emph{is} a Lie subalgebra of the algebroid (\ref{IIB algebra}). Of course, the full symmetry algebra
(\ref{IIB algebra}) satisfies the Jacobi identity, but is not a Lie algebra. It will be shown in section 3.4.2 that the action of
$\widetilde{X}^{mn}$ on all potentials is trivial so that the non-trivial action of the antisymmetric tensor transformation is generated by
$\widetilde{X}^{mnp}$ alone. Adding in the three form flux $M_{mnp}{}^a$ and allowing the twisted torus to be non-semi-simple, the symmetry
algebra contains the Lie algebra
\begin{eqnarray}
\left[Z_m,Z_n\right]&=&-f_{mn}{}^pZ_p-M_{mnp}{}^aY_a{}^p-K_{mnpqt}\widetilde{X}^{pqt}
\nonumber\\
\left[\widetilde{X}^{mnp},Z_q\right]&=&3f_{qt}{}^{[m}\widetilde{X}^{np]t}
\nonumber\\
\left[Y_a{}^m,Z_n\right]&=&f_{np}{}^mY_a{}^p-3\epsilon_{ab}M_{npq}{}^b\widetilde{X}^{mpq}
\end{eqnarray}
where $\widetilde{X}^{mnp}$ satisfies
\begin{equation}
O_{mnp}^{qt}\widetilde{X}^{mnp}=0   \qquad M_{mnp}{}^a\widetilde{X}^{mnp}=0
\end{equation}
These two constraints are required in order for the Jacobi identity to be satisfied. The first is simply a generalisation of the decomposition
in (\ref{decomposition}) for non-semi-simple $G$. To understand the second constraint $M_{mnp}{}^a\widetilde{X}^{mnp}=0$ consider the
transformation of the scalar field
\begin{equation}
\delta_X(\widehat{\Lambda})S_{(0)mnpq}=-O_{mnpq}^{tsl}\Lambda_{(0)tsl}
\end{equation}
and now using the decomposition
\begin{equation}
\Lambda_{(0)mnp}=\widetilde{\Lambda}_{(0)mnp}+M_{mnp}{}^a\widetilde{\Lambda}_{(0)a}
\end{equation}
Under the anti-symmetric tensor transformation generated by the parameter $M_{mnp}{}^a\widetilde{\Lambda}_{(0)a}$ the scalars $S_{(0)mnpq}$ are
invariant
\begin{eqnarray}
\delta_X(\widehat{\Lambda})S_{(0)mnpq}&=&-O_{mnpq}^{tsl}M_{tsl}{}^a\widetilde{\Lambda}_{(0)a}
\nonumber\\
&=&-6f_{[mn}{}^lM_{pq]l}{}^a\widetilde{\Lambda}_{(0)a}=0
\end{eqnarray}
where the last equality is a consequence of the flux integrability condition (\ref{flux constraints}). Therefore the symmetry with parameters
$\widetilde{\Lambda}_{(0)a}$ and $\widetilde{\Lambda}_{(0)mn}$ leave the scalar field $S_{(0)mnpq}$ invariant and drop out of the symmetry
algebra altogether, in accordance with the Jacobi identity above.

\subsection{Examples and Symmetry Breaking}

In this section the symmetry breaking down to a linearly realised subgroup that is generic for any solution shall be discussed. For vacua with
vanishing scalar expectation value, this is the complete breaking, but for non-trivial scalar expectation values there will be further breaking
through the standard Higgs mechanism. The examples considered in 3.4.2 and 3.4.3 are the two extreme cases, one where the fluxes vanish and the
reduction is the standard Scherk-Schwarz one and the second in which the structure constants $f_{mn}{}^p$ vanish but the fluxes do not. In the
following sub-section symmetry breaking in the gravity sector, i.e. the sector described by (\ref{s-s lagrangian}) is discussed. This symmetry
breaking is generic for any such theory.

\subsubsection{Symmetry Breaking in the Gravity Sector}

The breaking of the local $G_R$ symmetry by a choice of vacuum is easy to analyse in the Scherk-Schwarz reduction. The metric transforms in the
bi-adjoint representation
\begin{equation}
\delta_Z(\omega)g_{mn}=g_{mp}f_{nq}{}^p\omega^q+g_{np}f_{mq}{}^p\omega^q
\end{equation}
These transformations are only isometries for the cases where the metric is invariant $\delta_Z(\omega)g_{mn}=0$, i.e. the frame directions
$\sigma^{\bar{q}}$ for which
\begin{equation}
g_{mp}f_{n\bar{q}}{}^p+g_{np}f_{m\bar{q}}{}^p=0
\end{equation}
are isometric and the generators $Z_{\bar{q}}$ generate isometries of the metric $g_{mn}$ mediated by the gauge bosons $A^{\bar{q}}$. All
directions $\sigma^{\dot{q}}$ for which
\begin{equation}
g_{mp}f_{n\dot{q}}{}^p+g_{np}f_{m\dot{q}}{}^p\neq 0
\end{equation}
correspond to symmetries $Z_{\dot{q}}$ which are broken by the choice of vacuum. The gauge bosons $A^{\dot{q}}$ of these broken symmetries have
mass-like terms in the Lagrangian arising from the kinetic term $*Dg_{mn}\wedge Dg_{pq}$ of (\ref{s-s lagrangian})
\begin{equation} {\cal
L}_D=-\left(g^{mn}g^{pq}f_{m\dot{t}}{}^pf_{n\dot{s}}{}^q-f_{\dot{t}m}{}^nf_{\dot{s}n}{}^m\right)*A^{\dot{t}}\wedge A^{\dot{s}}+...
\end{equation}
If the metric $g_{mn}$ acquires a vacuum expectation value $\bar{g}_{mn}$, then this becomes a mass term for those graviphotons which are
\emph{not} associated to isometries of the frozen metric $\bar{g}_{mn}$, through the Higgs mechanism
\begin{equation} {\cal
L}_D=-\frac{1}{2}M_{\dot{t}\dot{s}}*A^{\dot{t}}\wedge A^{\dot{s}}+...
\end{equation}
where the mass matrix $M_{\dot{t}\dot{s}}$ is given by
\begin{equation}\label{graviphoton mass matrix}
M_{\dot{t}\dot{s}}=2\left(\bar{g}^{mn}\bar{g}^{pq}f_{m\dot{t}}{}^pf_{n\dot{s}}{}^q-f_{\dot{t}m}{}^nf_{\dot{s}n}{}^m\right)
\end{equation}
A vacuum in which the scalars have the expectation value $\bar{g}_{mn}=\eta_{mn}$, the (bi-invariant) Cartan-Killing metric
(\ref{cartan-killing}) will be invariant under $G_R$ while any other expectation value $\bar{g}_{mn}$ will break the gauge symmetry to the
subgroup preserving $\bar{g}_{mn}$.

\subsubsection{Reduction on an Identified Group Manifold with Semi-Simple Right Action}

The reduction on $\cX=G/\G$ where $G$ is any semi-simple group is considered. All fluxes are taken to be zero and therefore $\widehat{\cal
C}_{(4)}=\widehat{S}_{(4)}$ and $\widehat{\cal B}^a_{(2)}=\widehat{B}^a_{(2)}$. The gauge algebra in this case is a Lie algebra
\begin{eqnarray}
\left[Z_m,Z_n\right]&=&-f_{mn}{}^pZ_p
\nonumber\\
\left[X^{mnp},Z_q\right]&=&3f_{qt}{}^{[m}X^{np]t}
\nonumber\\
\left[Y_a{}^m,Z_n\right]&=&f_{np}{}^mY_a{}^p
\end{eqnarray}
with all other commutators vanishing. This algebra generates the group $G_R\ltimes U(1)^q$ where $q=d+{d\choose 3}$ which is broken to the
linearly realised subgroup $G_R\times U(1)^q$ by any vacuum of the theory as will be shown.

For the purposes of this section, the one-form antisymmetric tensor transformations are chosen to be
\begin{equation}\label{rhs}
\delta_Y(\widehat{\lambda})S_{(4)}=-\frac{1}{2}\epsilon_{ab}\widehat{\lambda}^a_{(1)}\wedge\widehat{H}^b_{(3)}
\end{equation}
where $\widehat{\cal H}^a_{(3)}=\widehat{H}^a_{(3)}$. This choice is related to that in (\ref{one form gauge transformation}) by a choice of the
arbitrary parameter $\Lambda_{(3)}$. The non-linear gauge transformations are
\begin{eqnarray}\label{broken symmetries}
\delta B^a_{(0)mn}&=&-\lambda^a_{(0)p}f_{mn}{}^p+...\nonumber\\
\delta S_{(2)mn}&=&-\Lambda_{(2)p}f_{mn}{}^p+...
\nonumber\\
\delta S_{(1)mnp}&=&3\Lambda_{(1)[m|t}f_{np]}{}^t+...
\nonumber\\
\delta S_{(0)mnpq}&=&-6\Lambda_{(0)[mn|t}f_{pq]}{}^t+...
\end{eqnarray}
where $+...$ denote linear terms. On a semi-simple identified group manifold the Cartan-Killing metric
\begin{equation}\label{cartan-killing}
\eta_{mn}=\frac{1}{2}f_{mp}{}^qf_{nq}{}^p
\end{equation}
is non-degenerate and invertible. The inverse metric $\eta^{mn}$ may be used to define $f_m{}^{np}=\eta^{nq}f_{mq}{}^p$. These constants may be
viewed as maps $f_{mn}{}^p\xi_p\rightarrow \xi_{mn}$ and $f_p{}^{mn}\xi_{mn}\rightarrow \xi_p$ for some antisymmetric
$\xi_{mn...}=\xi_{[mn...]}$
\begin{equation}
f:\R^d\rightarrow \R^{d\choose 2}    \qquad  f^{-1}:\R^{d\choose 2}\rightarrow \R^d
\end{equation}
and satisfy $f_{mp}{}^qf_q{}^{np}=2\delta_m{}^n$. It will also be useful to recall the definition of $O_{mnp}^{qt}$ and also to define the
constant $\Pi_{qt}^{mnp}$ as
\begin{eqnarray}
O_{mnp}^{qt}&=&3\delta_{[m}^qf_{np]}{}^t\nonumber\\
\Pi_{qt}^{mnp}&=&\frac{1}{2}\delta^{[m}_qf_t{}^{np]}
\end{eqnarray}
These constants may be viewed as maps; $\xi_{mnp}\rightarrow O_{mnp}{}^{[qt]}\xi_{qt}$ and $\xi_{mn}\rightarrow \Pi_{[mn]}{}^{pqt}\xi_{pqt}$, or
more abstractly as
\begin{equation}
O:\R^{d\choose 2}\rightarrow \R^{d\choose 3}    \qquad  \Pi:\R^{d\choose 3}\rightarrow \R^{d\choose 2}
\end{equation}
Note that these maps are not inverses of each other but satisfy the identity
\begin{equation}
\Pi_{[mn]}{}^{tsl}O_{tsl}{}^{[pq]}=\delta_{mn}{}^{pq}-\frac{1}{2}f_{mn}{}^tf_t{}^{pq}
\end{equation}
We also define constants $O_{mnpq}{}^{ts,l}$ and $\Pi^{mnpq}{}_{ts,l}$ as
\begin{eqnarray}
O_{mnpq}{}^{ts,l}&=&6\delta_{[mn}{}^{ts}f_{pq]}{}^l\nonumber\\
\Pi_{ts,l}{}^{mnpq}&=&\frac{1}{2}\delta_{ts}{}^{[mn}f_l{}^{pq]}
\end{eqnarray}
which may be thought of as maps defined by $O_{mnpq}{}^{[ts,l]}\xi_{tsl}=\xi_{mnpq}$ and $\Pi^{mnpq}{}_{[ts,l]}\xi^{tsl}=\xi^{mnpq}$
\begin{equation}
O:\R^{d\choose 3}\rightarrow \R^{d\choose 4}    \qquad  \Pi:\R^{d\choose 4}\rightarrow \R^{d\choose 3}
\end{equation}
and satisfy the relationship
\begin{equation}
\Pi_{[mn,p]}{}^{ijkl}O_{ijkl}{}^{[qt,s]}=\delta_{mnp}{}^{qts}-O_{mnp}{}^{[ij]}\Pi_{[ij]}{}^{qts}
\end{equation}
A number of other useful identities that these constants obey are collected in Appendix A of \cite{Hull:2006tp}. The following identities are
also useful
\begin{eqnarray}
D^2\xi_{m}&=&f_{mn}{}^p\xi_pF^n\nonumber\\
D^2\xi_{mn}&=&\left(O_{mnp}{}^{qt}-\delta_p{}^qf_{mn}{}^t\right)\xi_{qt}F^p
\end{eqnarray}
The potentials $S_{(0)mnpq}$, $S_{(1)mnp}$, $S_{(2)mn}$ and $S_{(3)m}$ take values in the $d \choose 4$, $d \choose 3$, $d \choose 2$ and $d$
dimensional representations of $SL(d;\R)$ respectively. In order to understand the possible field redefinitions that are required to remove the
non-linear group actions, the gauge parameters must be decomposed into these representations. The $\Lambda_{(2)m}$ can only take values in the
$d$ representation but we may decompose $\Lambda_{(1)mn}$ and $\Lambda_{(0)mnp}$ as
\begin{equation}
\Lambda_{(1)mn}=\widetilde{\Lambda}_{(1)mn}+f_{mn}{}^p\widetilde{\Lambda}_{(1)p}
\end{equation}
and
\begin{equation}\label{parameter decomposition}
\Lambda_{(0)mnp}=\widetilde{\Lambda}_{(0)mnp}+O_{mnp}{}^{qt}\widetilde{\Lambda}_{(0)qt}
\end{equation}
where $\widetilde{\Lambda}_{(1)mn}$ and $\widetilde{\Lambda}_{(0)mnp}$ satisfy\footnote{As discussed in a footnote to section 3.3.4, a further
decomposition $\widetilde{\Lambda}_{(0)mn}=\overline{\Lambda}_{(0)mn}+f_{mn}{}^p\overline{\Lambda}_{(0)p}$ (where
$f_p{}^{mn}\overline{\Lambda}_{(0)mn}=0$) is redundant since the parameter $\overline{\Lambda}_{(0)p}$ is projected out in (\ref{parameter
decomposition}) due to the identity $O_{mnp}{}^{qt}f_{qt}{}^s=0$. We may therefore neglect $\overline{\Lambda}_{(0)p}$ and enforce the
constraint $f_p{}^{mn}\widetilde{\Lambda}_{(0)mn}=0$ without loss of generality.}
\begin{equation}
f_p{}^{mn}\widetilde{\Lambda}_{(1)mn}=0 \qquad  \Pi_{qt}{}^{mnp}\widetilde{\Lambda}_{(0)mnp}=0
\end{equation}
In terms of these parameters the transformations (\ref{X-transformations}) are
\begin{eqnarray}
\delta_X(\Lambda)S_{(4)}&=&d\Lambda_{(3)}-\Lambda_{(2)m}\wedge F^m\nonumber\\
\delta_X(\Lambda)S_{(3)m}&=&D\left(\Lambda_{(2)m}+D\widetilde{\Lambda}_{(1)m}+\widetilde{\Lambda}_{(0)mn}F^n\right)
-\left(D\widetilde{\Lambda}_{(1)mn}-\widetilde{\Lambda}_{(1)mn}\right)\wedge F^n\nonumber\\
\delta_X(\Lambda)S_{(2)mn}&=&D\left(\widetilde{\Lambda}_{(1)mn}-D\widetilde{\Lambda}_{(0)mn}\right)
-\widetilde{\Lambda}_{(0)mnp}F^p\nonumber\\
&&-f_{mn}{}^p\left(\Lambda_{(2)p}+D\widetilde{\Lambda}_{(1)p}+\widetilde{\Lambda}_{(0)pq}F^q\right)\nonumber\\
\delta_X(\Lambda)S_{(1)mnp}&=&-D\widetilde{\Lambda}_{(0)mnp}+O_{mnp}^{qt}\left(D\widetilde{\Lambda}_{(1)qt}-\widetilde{\Lambda}_{(1)qt}\right)\nonumber\\
\delta_X(\Lambda)S_{(0)mnpq}&=&-O_{mnpq}^{tsl}\widetilde{\Lambda}_{(0)tsl}
\end{eqnarray}
The goldstone bosons of the broken symmetries (\ref{broken symmetries}) are given by
\begin{eqnarray}
\chi_{(0)m}^a&=&\frac{1}{2}f_m{}^{np}B^a_{(0)np}\nonumber\\
\chi_{(0)mnp}&=&\Pi_{mn,p}{}^{qtsl}S_{(0)qtsl}\nonumber\\
\chi_{(1)mn}&=&\Pi_{mn}{}^{pqt}S_{(1)pqt}\nonumber\\
\chi_{(2)m}&=&\frac{1}{2}f_m{}^{np}S_{(2)np}
\end{eqnarray}
These Goldstone bosons transform as
\begin{eqnarray}
\delta_Y(\lambda^a)\chi^a_{(0)m}&=&\lambda^a_{(0)m}\nonumber\\
\delta_X(\Lambda)\chi_{(0)mnp}&=&-\widetilde{\Lambda}_{(0)mnp}\nonumber\\
\delta_X(\Lambda)\chi_{(1)mn}&=&D\widetilde{\Lambda}_{(0)mn}-\widetilde{\Lambda}_{(1)mn}\nonumber\\
\delta_X(\Lambda)\chi_{(2)m}&=&\Lambda_{(2)m}+D\widetilde{\Lambda}_{(1)m}+\widetilde{\Lambda}_{(0)mn}F^n
\end{eqnarray}
One may therefore define the following $\delta_Y(\lambda^a_{(0)})$ and $\delta_X(\Lambda_{(0)mnp},\Lambda_{(1)mn},\Lambda_{(2)m})$-invariant
potentials
\begin{eqnarray}
\breve{S}_{(4)}&=&S_{(4)}-\chi_{(2)m}\wedge F^m\nonumber\\
\breve{S}_{(3)m}&=&S_{(3)m}-\frac{1}{2}\epsilon_{ab}\chi^a_{(0)m}H^b_{(3)}-D\chi_{(2)m}+\chi_{(1)mn}\wedge F^n\nonumber\\
\breve{S}_{(2)mn}&=&S_{(2)mn}+\epsilon_{ab}\chi^a_{(0)m}H^b_{(2)n}-D\chi_{(1)mn}+f_{mn}{}^p\chi_{(2)p}+\chi_{(0)mnp}F^p\nonumber\\
\breve{S}_{(1)mnp}&=&S_{(1)mnp}-\frac{3}{2}\epsilon_{ab}\chi^a_{(0)m}H^b_{(1)np}-D\chi_{(0)mnp}+O_{mnp}^{qt}\chi_{(1)qt}\nonumber\\
\breve{S}_{(0)mnpq}&=&S_{(0)mnpq}+2\epsilon_{ab}\chi^a_{(0)m}H^b_{(0)npq}-O_{mnpq}^{tsl}\chi_{(0)tsl}
\end{eqnarray}
similarly for the $B^a$-fields
\begin{eqnarray}
\breve{B}_{(2)}^a&=&B^a_{(2)}+\chi_{(0)m}^a\wedge F^m\nonumber\\
\breve{B}_{(1)m}^a&=&B^a_{(1)m}-D\chi_{(0)m}^a\nonumber\\
\breve{B}_{(0)mn}^a&=&B^a_{(0)mn}+f_{mn}{}^p\chi_{(0)p}^a
\end{eqnarray}
These field redefinitions take the form of infinitesimal gauge transformations (even though the Goldstone field need not be small) so the form
of the field strengths are not changed by the redefinition except to replace the potentials $(B^a,S)$ by the $(\breve{B}^a,\breve{S})$ defined
above. The gauge algebra is reduced to
\begin{equation}
[Z_m,Z_n]=-f_{mn}{}^pZ_p
\end{equation}
with all other commutators vanishing, generating the group $G_R\times U(1)^q$ as claimed.

\subsubsection{Flux Reduction on a Torus}

If $f_{mn}{}^p=0$, then the group $G_R$ is abelian and the internal manifold (after discrete identifications to compactify, if necessary) is a
torus and one may take $G _R = U(1)^d$. With flux ${\cal K}_{(5)}$ and $M^a_{(3)}$, the gauge algebra (\ref{IIB algebra}) has the Lie
sub-algebra
\begin{eqnarray}
\left[Z_m,Z_n\right]&=&-M_{mnp}{}^aY^p_a-K_{mnpqt}X^{pqt}\nonumber\\
\left[Y^m_a,Z_n\right]&=&-3\epsilon_{ab}M_{npq}{}^bX^{mpq}
\end{eqnarray}
with all other commutators vanishing.

\textbf{Non-linear Realisation of the Right Action}

As a warm up, consider the two-form sector in isolation. Viewing $\alpha^m \to \alpha^m M_{mnp}{}^a$ as a map
\begin{equation}
M:\R^d\rightarrow\R^{d(d-1)}
\end{equation}
the internal index $m$ can be split into $(m',\bar{m})$, so that $\bar{m}$ labels the $(d-d')$ dimensional kernel of
 the map $M$, and $m'$ labels the cokernel, so that
\begin{equation}
M_{mn\bar{p}}{}^a =0 \qquad M_{m'n'p'}{}^a \neq 0
\end{equation}
Then the  transformation of the $B^a_{(0)}$ scalars is
\begin{equation}
\delta B^a_{(0){ n'p'}} =\omega^{m'}M_{m'n'p'}{}^a , \qquad \delta B^a_{(0){ m\bar{n}}} = 0
\end{equation}
The transformations generated by $Z_{m'}$ with parameters $\omega^{m'}$ are spontaneously broken by any vacuum of the theory. For a vacuum
$(\bar{g},\bar{\phi},\bar{{\cal K}})$ the $A^{m'}$ fields have mass term in the Lagrangian
\begin{eqnarray}
\mathcal{L}_D=-\frac{1}{2}e^{-4\beta\bar{\phi}}\bar{g}^{mn}\bar{g}^{pq}\bar{{\cal K}}_{ab}M_{mpt'}{}^aM_{nqs'}{}^b*A^{t'}\wedge A^{s'}+...
\end{eqnarray}
The $2d$ dimensional gauge group is broken to the $(2d-d')$ dimensional abelian subgroup $U(1)^{2d-d'}$ generated by $Z_{\bar m}$ and $X^m$ with
parameters $\omega^{\bar{m}}$ and $\lambda_{(0)m}$ respectively. Let $\widetilde M^{m'n'p'}{}_a$  be any constants satisfying $\widetilde
M^{m'n'p'}{}_aM^b_{ n'p'q'}=\delta ^{m'}{}_{q'}\delta _a{}^b$. Then the Goldstone fields $\chi_{(0)}{}^{m'}$ are defined by
\begin{equation}
    \chi_{(0)}{}^{m'}=\tilde M^{m'n'p'}{}_aB^a_{(0)n'p'}
\end{equation}
transforming as a shift
\begin{equation}
\delta B^a_{(0)\bar{M}}=0 \qquad \delta\chi_{(0)}{}^{m'}=\omega^{m'}
\end{equation}
The remaining scalars are invariant, $\delta B^a_{(0){ m\bar{n}}} = 0$. The massive graviphotons are defined as
$\breve{A}^{m'}=A^{m'}+d\chi_{(0)}{}^{m'}$ and are singlets of the gauge transformations.

\textbf{Non-linear Realisation of the One-Form Antisymmetric Tensor Transformation}

If the four form potentials are now introduced, the one-form antisymmetric tensor transformations appear as shift symmetries
\begin{eqnarray}\label{shifts}
\delta B^a_{(0)mn}&=&M_{mnp}{}^a\omega^p+...\nonumber\\
\delta S_{(1)mnp}&=&-\epsilon_{ab}\lambda^a_{(1)}M_{mnp}{}^b+...\nonumber\\
\delta S_{(0)mnpq} &=&- 4\epsilon_{ab} \lambda^a_{(0)m}M_{npq}{}^b+K_{mnpqt}\omega^t+...
\end{eqnarray}
Therefore, even on a flat torus, the presence of flux will break some of the anti-symmetric tensor transformations generated by $Y^m_a$. The
gauge bosons of symmetries with parameters $\lambda_{(0)m}^a$ and $\lambda^a_{(1)}$ are $B^a_{(2)}$ and $B^a_{(1)m}$ respectively and some of
these potentials become massive by the Higgs mechanism. As in the flat torus case above, one may define constants $\widetilde{M}^{mnp}{}_a$ such
that $M_{mnp}{}^b\widetilde{M}^{mnp}{}_a=\delta_a{}^b$, such that the goldstone one-form $\chi_{(1)a}$ defined as
\begin{equation}
\chi_{(1)a}=S_{(1)mnp}\widetilde{M}^{mnp}{}_a
\end{equation}
which transforms as $\delta \chi_{(1)a}=\epsilon_{ab}\lambda^b_{(1)}+...$. The $S_{(0)mnpq}$ transformation in (\ref{shifts}) may be written as
\begin{eqnarray}
\delta S_{(0)mnpq} &=&- 4\epsilon_{ab} \lambda^a_{(0)m}M^b_{npq}+K_{mnpqt}\omega^t+...\nonumber\\
&=&-R_{mnpq}{}^t{}_a\lambda^a_{(0)t}+K_{mnpqt}\omega^t+...
\end{eqnarray}
where it is useful to define the constant $R_{mnpq}{}^t{}_a=4\epsilon_{ab}\delta^t{}_{[m}M^b_{npq]}$. This transformation may be written as
\begin{eqnarray}
\delta S_{(0)\Sigma}&=&\left(%
\begin{array}{cc}
  -\lambda_{(0)t}^a & \omega^t \\
\end{array}%
\right)\left(%
\begin{array}{c}
  R_{mnpq}{}^t{}_a \\
  K_{mnpqt} \\
\end{array}%
\right)+...\nonumber\\ &=&\alpha_{(0)M}t^M{}_{\Sigma}+...
\end{eqnarray}
where the index $M=1,2...3d$ so that
\begin{equation}
\alpha_{(0)M}=\left(%
\begin{array}{ccc}
  -\lambda_{(0)m}^1 & -\lambda_{(0)m}^2 & \omega^m \\
\end{array}%
\right)
\end{equation}
and the index $\Sigma=[mnpq]=1,2...{d\choose 4}$. Treating $t^M{}_{\Sigma}$ as the map
\begin{equation}
t:\R^{3d}\rightarrow\R^{d\choose 4}
\end{equation}
the index $M$ may be split into $M=(M',\bar{M})$ where $M'$ and $\bar{M}$ label the cokernel and kernel of the map $t$ respectively. A basis may
then be chosen such that the constant tensor $t^M{}_{\Sigma}$ is written as
\begin{equation}
t^M{}_{\Sigma}=\left(%
\begin{array}{cc}
  t^{M'}{}_{\Sigma'} & 0 \\
  0 & 0 \\
\end{array}%
\right)
\end{equation}
 The choice of basis is such that there exists an inverse $\widetilde{t}^{\Sigma'}{}_{M'}$ where
$t^{M'}{}_{\Sigma'}\widetilde{t}^{\Sigma'}{}_{N'}=\delta^{M'}{}_{N'}$ and
$\widetilde{t}^{\Sigma'}{}_{M'}t^{M'}{}_{\Lambda'}=\delta^{\Sigma'}{}_{\Lambda'}$. The Goldstone boson for the symmetry with parameter
$\alpha_{(0)M'}$ is
\begin{eqnarray}
\chi_{(0)M'}=\widetilde{t}^{\Sigma'}{}_{M'}S_{(0)\Sigma'}
\end{eqnarray}
It is useful to combine the generators $Z_m$ and $Y_a^m$ into the doublet
\begin{equation}
T^{M}=\left(%
\begin{array}{c}
  Y^m_a \\
  Z_m \\
\end{array}%
\right)
\end{equation}
so that $\delta=\alpha_MT^M$. Those symmetries generated by $T^{\bar{M}}$ (with parameter $\alpha_{\bar{M}}$) are preserved whilst those
generated by $T^{M'}$ (with parameter $\alpha_{M'}$) have non-linear realisations and are always broken by a choice of vacuum of the theory.
Gauge singlet fields may be defined
\begin{eqnarray}
\breve{S}_{(0)\Sigma'}&=&S_{(0)\Sigma'}-\chi_{(0)M'}t^{M'}{}_{\Sigma'}\nonumber\\
\breve{S}_{(1)mnp}&=&S_{(1)mnp}+\chi_{(1)a}M_{mnp}{}^a\nonumber\\
\breve{B}^a_{(0)mn}&=&B_{(0)mn}-\chi_{(0)}{}^pM_{mnp}{}^a
\end{eqnarray}
Both the ${\cal H}^a_{(1)mn}$ and $G_{(1)mnpq}$ field strengths contribute to the graviphoton mass term where $G_{(1)mnpq}=K_{mnpqt}A^t+...$ and
${\cal H}^a_{(1)mn}=M_{mnp}{}^aA^p+...$. For a given vacuum expectation value of the scalars $\bar{g}$, $\bar{{\cal K}}$ and $\bar{\varphi}$,
the mass-like term in the Lagrangian due to these field strengths may be written as
\begin{eqnarray}
{\cal L}_D&=&-\frac{1}{2}{\cal M}_{AB}*{\cal A}_{(1)}^A\wedge {\cal A}_{(1)}^B+...
\end{eqnarray}
where the components of the mass matrix are
\begin{eqnarray}
{\cal
M}_{mn}&=&e^{4\beta\bar{\varphi}}\bar{g}^{mt}\bar{g}^{ns}\bar{g}^{pl}\bar{g}^{qk}\left(e^{4(\alpha-\beta)\bar{\varphi}}K_{mnpqi}K_{tslkj}+\frac{1}{d}\bar{g}_{lq}\bar{g}_{pk}\bar{{\cal
K}}_{ab}M_{mni}{}^aM_{tsj}{}^b\right)\nonumber\\
{\cal
M}_m{}^n&=&-2e^{4\beta\bar{\varphi}}\bar{g}^{mt}\bar{g}^{ns}\bar{g}^{pl}\bar{g}^{qk}e^{4(\alpha-\beta)\bar{\varphi}}\epsilon_{ab}M_{tsl}{}^aK_{mnpqi}\nonumber\\
{\cal M}^{mn}&=&4e^{4\beta\bar{\varphi}}\bar{g}^{mt}\bar{g}^{ns}\bar{g}^{pl}\bar{g}^{qk}\epsilon_{ac}\epsilon_{bd}M_{mnp}{}^cM_{qts}{}^d
\end{eqnarray}
and
\begin{equation}
{\cal A}_{(1)}^A=\left(%
\begin{array}{c}
  A^m \\
  B_{(1)m}^a \\
\end{array}%
\right)
\end{equation}
For a given vacuum expectation value of the scalar fields the diagonalisation of the matrix ${\cal M}_{AB}$ gives the (mass)$^2$ spectrum of the
${\cal A}^A_{(1)}$ potentials. In general, with non-trivial internal geometry some of the $S_{(1)mnp}$ gauge fields will become massive,
corresponding to the breaking of symmetries generated by $\widetilde{X}^{mnp}$. Upon symmetry breaking by the choice of some vacuum
$(\bar{g},\bar{\varphi},\bar{\cal K})$, the general effective theory will contain massive one-forms and massless gauge bosons that are linear
combinations of the $B^a_{(1)m}$, $A^m$ and $S_{(1)mnp}$.

\section{Compactifications with S-Duality Twists and F-Theory}

Consider a $D+d+1$ dimensional field theory coupled to gravity. The theory is reduced on a $d$-dimensional torus $T^d$, with real coordinates
$z^a\sim z^a+1$ where $a=1,2...d$. This produces a theory in $D+1$ dimensions with scalar fields that include those
 in the coset $GL(d,\R)/SO(d)$ arising from the torus
moduli. Truncating to the $z^a$ independent zero mode sector, this theory has a global symmetry $U$ that contains the $GL(d,\R)$ arising from
diffeomorphisms of the $d$-dimensional torus. In the full Kaluza-Klein theory this is broken to the $GL(d,\Z)$ that acts as large
diffeomorphisms on the $d$-dimensional torus similarly, in string theory $U$ is broken to the discrete U-duality subgroup $U(\Z)$. The action of
$U$ on fields $\psi$ of the reduced theory in some representation of $U$ is denoted as $\psi\rightarrow \gamma[\psi]$.

The duality twist reductions of this theory describes reduction on a further circle with periodic coordinate $y\sim y+1$, twisting the fields
over the circle by an element of $U$ using the ansatz \cite{Hull:2003kr,Hull:1998vy,Hull:2002wg,Dabholkar:2002sy}
 \begin{equation}\label{s-s ansatz}
\psi(x^\m,y)=\gamma_y[\psi(x^\m)]
\end{equation}
where $x^\m$ are the $D$ non-compact spacetime coordinates. Consistency of the reduction, in the sense described in the introduction, requires
the reduced theory to be independent of $y$, which is achieved by choosing the form of $\gamma$ to be
\begin{equation}\label{twist}
\gamma(y)=exp\left(My\right)
\end{equation}
for some matrix $M$ in the Lie algebra of $U$.

The map $\gamma(y)$ is not periodic, but has monodromy $\mathcal{M}(\gamma)=\gamma(0)\gamma(1)^{-1}=e^M$ in $U$ and the physically distinct
reductions are classified by the conjugacy class of the monodromy  \cite{Hull:1998vy}. In the full theory in which all massive modes are kept,
$U$ is typically broken to a discrete subgroup $U(\Z)$. In order for $\Psi (x,y+2\pi)={\cal M}\Psi (x,y) $ to be well-defined, the monodromy
${\cal M}$ must therefore be in the symmetry group $U(\Z)$ \cite{Dabholkar:2002sy,Hull:1994ys}.

If the monodromy is in the geometric $GL(d;\Z)$ sub-group then the reduction may be viewed as a specific class of the identified group manifold
reductions of the previous sections. In this case the identified group manifold really is a topologically twisted torus fibration. If the
monodromy is more general then the reduction cannot be given as a purely geometric construction as the monodromy (or transition functions
between patches) may now be S or T dualities. For example, if the monodromy is an element of the T-duality group, then the string theory is only
defined in a certain patch and we must consider the identification of a particular string theory as only possible locally. Globally, this
picture must be generalised to include the string theory and its T-dual. In particular, the transition function would invert the radii of the
circles on a torus and generate non-trivial $B$-fields - clearly a non-geometric operation. The introduction of Ramond fields and their
associated D-branes on a background with a monodromy in the factorised duality subgroup of $O(d,d)$ leads to further startling features
\cite{Hull:2003kr,Shelton:2005cf}. For example, the dimension of the D-brane would not be globally defined as the transition function increases
or decreases the dimension of the brane as one moves between patches \cite{Shelton:2005cf,Hull:2004in}. Another example is that of string theory
on a Calabi-Yau manifold \cite{Dabholkar:2002sy}. Here the backgrounds are permitted to have transition functions which are mirror symmetries.
It should be noted that truncation to the lowest modes on a Calabi-Yau is not consistent so the detailed analysis in that case is expected to be
more complicated.

The situation is even more drastic if the monodromy is an element of S-duality. In this case, the perturbative string theory picture can only be
used locally. In this section such reductions of IIB supergravity with S-duality twists are investigated at the supergravity level. Such
reductions have also been investigated in \cite{Hull:2003kr,Cowdall:2000sq}.

\subsection{Reductions with S-Duality Twists}

S-duality is non-perturbative in the string coupling $g_s$, mixing Ramond and Neveu-Schwarz fields, and therefore cannot be given a worldsheet
interpretation but there is compelling evidence to believe that this symmetry is an exact symmetry of the full non-perturbative theory
\cite{Sen:1998kr}.

The reduction to nine dimensions of the bosonic sector of IIB supergravity on a circle with an S-duality twist will be considered. The
ten-dimensional IIB Lagrangian, written in a manifestly $SL(2)$ invariant form was given by (\ref{IIB Lagrangian}). Consider the reduction
ansatz
\begin{equation}\label{line element}
ds^2_{D+1}=e^{2\alpha\varphi}ds^2_D+e^{2\beta\varphi}(dy+A)^2
\end{equation}
where $y$ parameterises the circle direction. The duality twist ansatz (\ref{s duality}), (\ref{s-s ansatz}) and (\ref{twist}) to reduce the
fields of the IIB theory on circle with S-duality twist are
\begin{equation}\label{Twisted ansatz}
\widehat{{\cal K}}^a{}_b(x,y)=e^{(M^t)^a{}_cy}{\cal K}^c{}_d(x)e^{M^d{}_by} \qquad
\widehat{B}^a_{(2)}(x,y)=e^{M^a{}_by}\left(B^b_{(2)}(x)+B^b_{(1)}(x)\wedge\nu\right)
\end{equation}
where $M$ is a twist matrix in the Lie algebra of $SL(2)$ and $M^t$ its transpose. The reduced scalar Lagrangian is then\footnote{We have made
use of the fact that $Tr(M^t\cdot M^t)=Tr(M\cdot M)$ in the potential.}
\begin{equation}
{\cal L}_{\cal K}=\frac{1}{4}Tr\left(*D{\cal K}\wedge D{\cal K}^{-1}\right)-\frac{1}{2}e^{-2(\alpha+\beta)\varphi}Tr\left(M^2+M^T{\cal K}M{\cal
K}^{-1}\right)*1
\end{equation}
where the covariant derivatives are
\begin{eqnarray}
D_i{\cal K}^a{}_b&=&\partial_i{\cal K}^a{}_b-(M^T{\cal K}+{\cal K}M)^a{}_bA_i  \nonumber\\ D_i({\cal K}^{-1})_a{}^b&=&\partial_i({\cal
K}^{-1})_a{}^b+({\cal K}^{-1}M^T+M{\cal K}^{-1})_a{}^bA_i
\end{eqnarray}
The reduction of the two form field strength term
\begin{equation}
{\cal L}_{\widehat{H}}=-\frac{1}{2}{\cal K}_{a b}*\widehat{H}^{a}_{(3)} \wedge \widehat{H}^{b}_{(3)}
\end{equation}
gives the low energy effective term
\begin{equation}
{\cal L}_H=-\frac{1}{2}{\cal K}_{a b}e^{-4\alpha\varphi}*H^{a}_{(3)} \wedge H^{b}_{(3)}-\frac{1}{2}{\cal K}_{a
b}e^{-2(\alpha+\beta)\varphi}*H^{a}_{(2)} \wedge H^{b}_{(2)}
\end{equation}
where the reduced field strengths are
\begin{eqnarray}
H^a_{(3)}&=&dB^a_{(2)}+M^a{}_bB_{(2)}^b\wedge A-B^a_{(1)}\wedge F\nonumber\\
&=&DB^a_{(2)}-B_{(1)}^a\wedge F\nonumber\\
H_{(2)}^a&=&dB^a_{(1)}-M^a{}_bB_{(1)}^b\wedge A-M^a{}_bB_{(2)}^b\nonumber\\
&=&DB_{(1)}^a-M^a{}_bB_{(2)}^b\nonumber\\
F&=&dA
\end{eqnarray}
with the Bianchi identites
\begin{equation}
DH_{(3)}^a=0  \qquad  DH_{(2)}^a=0  \qquad  dF=0
\end{equation}
The self dual five form field strength, although a singlet under the S-duality transformation, still has a non-trivial deformation in the
Scherk-Schwarz ansatz coming from the Chern-Simons terms. The five form term in the action is
\begin{equation}
{\cal L}_{\widehat{G}}=-\frac{1}{4}*\widehat{G}_{(5)} \wedge \widehat{G}_{(5)}
\end{equation}
which reduces to
\begin{equation}
{\cal L}_G=-\frac{1}{4}e^{-8\alpha\varphi}*G_{(5)} \wedge G_{(5)}-\frac{1}{4}e^{-(2\alpha+\beta)\varphi}*G_{(4)} \wedge G_{(4)}
\end{equation}
where
\begin{eqnarray}
G_{(5)}&=&dC_{(4)}-C_{(3)}\wedge
F+\frac{1}{2}\epsilon_{ab}B^a_{(2)}\wedge H^b_{(3)}\nonumber\\
G_{(4)}&=&dC_{(3)}-\frac{1}{2}\epsilon_{ab}\left(B_{(1)}^a\wedge H^b_{(3)}-B^a_{(2)}\wedge H_{(2)}^b\right)
\end{eqnarray}
Finally, the Chern-Simons terms reduce to
\begin{equation}
{\cal L}_{cs}=-\frac{1}{4}\left(C_{(3)}\wedge H^a_{(3)}\wedge H^b_{(3)}+2C_{(4)}\wedge H^a_{(3)}\wedge H^b_{(2)}\right)
\end{equation}
The self duality constraint $\widehat{G}_{(5)}=*\widehat{G}_{(5)}$ reduces to a relationship between $\widehat{G}_{(5)}$ and $\widehat{G}_{(4)}$
and must be imposed on the equations of motion of the reduced theory.

\subsection{Gauge Symmetry and its Breaking}

In this section the symmetries of the reduced Lagrangian are studied and the symmetry breaking and mass mechanisms involved are analysed.

\subsubsection{Antisymmetric Tensor Transformations}

In ten dimensions the three from field strength $H^a_{(3)}$ is invariant under the abelian antisymmetric tensor transformation of the potential
$\delta \widehat{B}^a_{(2)}=d\widehat{\lambda}^a_{(1)}$. The four form potential transforms to compensate for the transformation of the
Chern-Simons term in the field strength $\widehat{G}_{(5)}$ as
\begin{equation}
\delta\widehat{C}_{(4)}=-\frac{1}{2}\epsilon_{ab}\widehat{\lambda}^a_{(1)}\wedge \widehat{H}^b_{(3)}
\end{equation}
The combined transformations of the four-form and two-form potentials leave the five-form field strength, $\widehat{G}_{(5)}$, invariant. Under
the S-duality twisted reduction considered in the preceding section, the reduced potentials transform as
\begin{eqnarray}\label{symmetry transformations}
\delta B_{(2)}^a&=&d\lambda^a_{(1)}-M^a{}_b\lambda_{(1)}^b\wedge
A+\lambda_{(0)}^aF\nonumber\\
\delta B_{(1)}^a&=&d\lambda^a_{(0)}+M^a{}_b\lambda_{(0)}^b\wedge
A+M^a{}_b\lambda_{(1)}^b\nonumber\\
\delta C_{(4)}&=&-\frac{1}{2}\epsilon_{ab}\lambda^a_{(1)}\wedge
H^b_{(3)}\nonumber\\
\delta C_{(3)}&=&-\frac{1}{2}\epsilon_{ab}\left(\lambda^a_{(1)}\wedge H^b_{(2)}-\lambda^a_{(0)}H^b_{(3)}\right)
\end{eqnarray}
The  transformation $\delta B_{(1)}^a=M^a{}_b\lambda_{(1)}^b+...$ is a shift symmetry, i.e. it is non-linear realisation of the symmetry group
and will not be preserved by any vacuum of the theory. A massive two-form $\breve{B}_{(2)}^a$ may be defined
\begin{equation}\label{massive form1}
\breve{B}_{(2)}^a=B_{(2)}^a-(M^{-1})^a{}_bDB_{(1)}^b
\end{equation}
where $B^a_{(2)}$ has eaten $B_{(1)}^a$ to become massive and $\breve{B}_{(2)}^a$ is a singlet of the symmetry transformations (\ref{symmetry
transformations}). The redefinition (\ref{massive form1}) is dependent on the existence of the inverse $(M^{-1})^a{}_b$. It will be shown that
that this is not always the case and that care must be taken in defining the massive two form fields (\ref{massive form1}). For now it will be
assumed that the mass matrix is chosen such that $(M^{-1})^a{}_b$ exists. Applying the field redefinitions
\begin{eqnarray}
\breve{C}_{(3)}&=&C_{(3)}+\frac{1}{2}\epsilon_{ab}(M^{-1})^a{}_cB_{(1)}^c\wedge
H_{(2)}^b\nonumber\\
\breve{C}_{(4)}&=&C_{(4)}+\frac{1}{2}\epsilon_{ab}(M^{-1})^a{}_cB_{(1)}^c\wedge H_{(3)}^b
\end{eqnarray}
the field strengths become
\begin{eqnarray}
G_{(5)}&=&d\breve{C}_{(4)}-\breve{C}_{(3)}\wedge F+\frac{1}{2}\epsilon_{ab}\breve{B}^a_{(2)}\wedge
D\breve{B}^b_{(2)}\nonumber\\
G_{(4)}&=&d\breve{C}_{(3)}-\frac{1}{2}M_{ab}\breve{B}^a_{(2)}\wedge
\breve{B}^b_{(2)}\nonumber\\
H_{(3)}^a&=&D\breve{B}^a_{(2)}\nonumber\\
H_{(2)}^a&=&-M^a{}_b\breve{B}^b_{(2)}
\end{eqnarray}
where $M_{ab}=\epsilon_{(a|c}M^c{}_{|b)}$. $B_{(1)}^a$ is eaten by $\breve{B}^a_{(2)}$ and completely drops out of all of the field equations.
The $\breve{C}_{(3)}$ and $\breve{C}_{(4)}$ fields remain massless and charged under the abelian gauge symmetry generated by the transformations
\begin{equation}
\delta \breve{C}_{(3)}=d\Lambda_{(2)}   \qquad  \delta \breve{C}_{(4)}=d\Lambda_{(3)}
\end{equation}
The Lagrangian of the reduced theory is then
\begin{eqnarray}\label{massive Lagrangian}
{\cal L}_9&=&R*1 - \frac{1}{2}*d\varphi\wedge d\varphi -\frac{1}{2}e^{2(\beta-\alpha)\varphi}*F \wedge F+\frac{1}{4}Tr\left(*D{\cal K}\wedge
D{\cal
K}^{-1}\right)\nonumber\\
&&-\frac{1}{4}e^{-8\alpha\varphi}*G_{(5)} \wedge G_{(5)}-\frac{1}{4}e^{-(2\alpha+\beta)\varphi}*G_{(4)} \wedge
G_{(4)}\nonumber\\
&&-\frac{1}{2}{\cal K}_{a b}e^{-4\alpha\varphi}*D\breve{B}^a_{(2)} \wedge D\breve{B}^b_{(2)}-\frac{1}{2}{\cal K}_{a
b}e^{-2(\alpha+\beta)\varphi}M^a{}_cM^b{}_d*\breve{B}^c_{(2)} \wedge \breve{B}^d_{(2)}
\nonumber\\
&&-\frac{1}{4}\epsilon_{ab}\left(\breve{C}_{(3)}\wedge H^a_{(3)}\wedge H^b_{(3)}+2\breve{C}_{(4)}\wedge H^a_{(3)}\wedge H^b_{(2)}\right)
\nonumber\\
&&-\frac{1}{2}e^{-2(\alpha+\beta)\varphi}Tr\left(M^2+M^t{\cal K}M{\cal K}^{-1}\right)*1
\end{eqnarray}

In some cases $M^a{}_b$ may not be invertible. An example is the reduction with parabolic twist where
\begin{equation}
M_p=\left(%
\begin{array}{cc}
  0 & m \\
  0 & 0 \\
\end{array}%
\right) \qquad  m\in\Z
\end{equation}
This mass matrix has no inverse so one must be careful in defining the massive two form (\ref{massive form1}). For such non-invertible matrices
one may always choose a basis such that the mass matrix takes the form
\begin{equation}
M=\left(%
\begin{array}{cc}
  M & 0 \\
  0 & 0 \\
\end{array}%
\right)
\end{equation}
In this basis the potentials are written as
\begin{equation}
B^a_{(1)}=\left(%
\begin{array}{c}
  B'_{(1)} \\
  \bar{B}_{(1)} \\
\end{array}%
\right) \qquad  B^a_{(2)}=\left(%
\begin{array}{c}
  B'_{(2)} \\
  \bar{B}_{(2)} \\
\end{array}%
\right)
\end{equation}
It is then possible to identify a massive two-form
\begin{equation}\label{massive form}
\breve{B}_{(2)}=B'_{(2)}-M^{-1}DB'_{(1)}
\end{equation}
whilst $\bar{B}_{(1)}$ and $\bar{B}_{(2)}$ remain massless.

\subsubsection{Internal Diffeomorphism and Fixed Points of the Twist}

In addition to the antisymmetric tensor transformations, the reduced theory has a $U(1)$ gauge symmetry originating from diffeomorphisms
$y\rightarrow y-\omega(x)$ along the compactification circle. The reduced fields transform as
\begin{eqnarray}
\delta A&=&d\omega\nonumber\\
\delta B^a_{(1)}&=&M^a{}_bB^b_{(1)}\omega\nonumber\\
\delta B^a_{(2)}&=&M^a{}_bB^b_{(2)}\omega\nonumber\\
\delta {\cal K}_{ab}&=&-2{\cal K}_{(a|c}M^c{}_{|b)}\omega
\end{eqnarray}
where $\omega=\omega(x)$. For a given expectation value of the scalars $\langle{\cal K}\rangle_{ab}=\overline{\cal K}_{ab}$ this symmetry will
be broken unless
\begin{equation}\label{Isometry}
\langle\delta {\cal K}\rangle_{ab}=-2\overline{{\cal K}}_{(a|c}M^c{}_{|b)}\omega=0
\end{equation}
If this is the case, then the graviphotons are massless, as may be seen from the $D{\cal K}$ term in the Lagrangian (\ref{massive Lagrangian})
\begin{equation}
{\cal L}_9=\frac{1}{4}Tr\left(*D\overline{{\cal K}}\wedge D\overline{{\cal K}}^{-1}\right)+...=-\frac{1}{2}\mu^2*A\wedge A+...
\end{equation}
where the mass $\mu$ is given by
\begin{equation}
\mu^2=Tr\left(M^2+M^t\overline{{\cal K}}M\overline{{\cal K}}^{-1}\right)
\end{equation}
This $\mu^2$ term is proportional to the scalar potential of the reduction and therefore the graviphotons will be massless at the minima of the
potential. The minima of the potential for such reductions was studied in \cite{Dabholkar:2002sy} where it was shown that the potential is zero
only for elliptic twists, an argument that is reviewed here. The moduli matrix may be written in terms of the $SL(2)/SO(2)$ zweibein ${\cal V}$
as ${\cal K}_{ab}={\cal V}_a{}^{\alpha}\delta_{\alpha\beta}{\cal V}^{\beta}{}_b$ where
\begin{equation}
{\cal V}=e^{\frac{1}{2}\phi}\left(%
\begin{array}{cc}
  1 & C_{(0)} \\
  0 & e^{-\phi} \\
\end{array}%
\right)
\end{equation}
and $\alpha,\beta=1,2$ are $SO(2)$ matrix representation indices. The vanishing point for the potential and graviphoton mass occurs when the
complex structure has the vacuum value $\langle{\cal V}\rangle={\cal V}_0$ such that the twist matrix $M$ is equivalent to a matrix $R$ in the
Lie algebra of $SO(2)$ up to a conjugation by ${\cal V}_0$
\begin{equation}
M={\cal V}^{-1}_0R{\cal V}_0
\end{equation}
This may be seen as follows. Setting $\overline{{\cal K}}={\cal V}^t_0{\cal V}_0$ and $M={\cal V}^{-1}_0R{\cal V}_0$ the potential may be
written as \cite{Dabholkar:2002sy}
\begin{equation}
V\propto\mu^2=\frac{1}{2}Tr\left(Y^2\right)
\end{equation}
where $Y=R+R^t$. Since $R$ is in the Lie algebra of $SO(2)$, $Y=0$ and the potential vanishes and the graviphotons are massless. As recognised
in \cite{Dabholkar:2002sy} this is to expected as such a choice of vacuum is a fixed point of the twist action and therefore will have no effect
on the field theory. For this choice of scalars (\ref{Isometry}) can be written
\begin{eqnarray}\label{vacuum variation}
\langle\delta{\cal K}\rangle_{ab}&=&-({\cal V}_0)_a{}^{\alpha}R_{\alpha\beta}({\cal V}_0)^{\beta}{}_b-({\cal V}_0)_b{}^{\alpha}R_{\alpha\beta}({\cal V}_0)^{\beta}{}_a\nonumber\\
&=&-2({\cal V}_0)_a{}^{\alpha}R_{(\alpha\beta)}({\cal V}_0)^{\beta}{}_b
\end{eqnarray}
The right hand side of (\ref{vacuum variation}) vanishes as the generator $R_{\alpha\beta}$ of $SO(2)$ is antisymmetric. Therefore a choice of
vacuum will generally break the $U(1)$ isometry group unless the twist is in the elliptic conjugacy class.

\subsection{Non-Geometric Twists and F-Theory}

F-Theory is formally a twelve dimensional theory which when reduced on $T^2$ gives a theory whose truncation to the massless sector gives the
IIB supergravity. The S-duality of the IIB theory is then described geometrically as the mapping class group of the $T^2$ fibre for which the
axio-dilaton $\tau=C_{(0)}+ie^{-\phi}$ is the complex structure. This, apparently redundant, description of the IIB theory becomes useful when
one considers compactifications of F-Theory on spaces that have a $T^2$ fibration that is not trivial \cite{Kumar:1996zx}. The relevance of this
picture here is that in some cases it may be used to give a geometrical interpretation to otherwise non-geometric duality twist backgrounds.

Consider for example the $SL(2,\Z)$ U-duality of the IIB string theory \cite{Hull:1994ys}. Reducing from 10 to 9 dimensions on a circle with
monodromy in $SL(2,\Z)$ investigated in the previous section and also \cite{Hull:1998vy,Hull:2002wg,Bergshoeff:2002mb,Meessen:1998qm}. As the
$SL(2,\Z)$ symmetry is not geometric, this cannot be realised as a compactification in the conventional sense. However, it can be realised as a
\lq compactification' of F-theory on the twisted torus corresponding to a $T^2$ bundle over $S^1$ with $SL(2,\Z)$ monodromy \cite{Hull:1998vy}.
For example, the case of an elliptic twist with vanishing potential discussed in the last section may be thought of as a reduction of F-Theory
on an orbifold \cite{Dabholkar:2002sy}.

This further extends the notion of a non-geometric background to a non-perturbative background where one must cover the internal circle with at
least two patches and where the transition functions between patches are S-dualities. If these Scherk-Schwarz reductions lift to solutions of
the full M-Theory one must accept that, even at weak coupling, perturbative string theory can at best describe only the local physics of such
solutions.

\begin{center}
\textbf{Acknowledgements}

The author would like to thank Chris Hull, Paul Townsend and Dan Waldram for useful discussions.
\end{center}

\newpage

\appendix

\section{Bianchi Identities and Field Strengths}

The reduced field strengths are
\begin{eqnarray}
{\cal H}^a_{(3)}&=&dB^a_{(2)}+B^a_{(1)m}\wedge F^m+\frac{1}{6}M^a_{mnp}A^m\wedge A^n\wedge A^p
\nonumber\\
{\cal H}_{(2)m}^a&=&DB^a_{(1)m}+B^a_{(0)mn}F^n+\frac{1}{2}M^a_{mnp}A^n\wedge A^p
\nonumber\\
{\cal H}^a_{(1)mn}&=&DB^a_{(0)mn}+f^p_{mn}B^a_{(1)p}+M^a_{mnp}A^p
\nonumber\\
{\cal H}^a_{(0)mnp}&=&3B^a_{(0)[m|q}f^q_{|np]}+M^a_{mnp}
\end{eqnarray}
and for the five form field strength
\begin{eqnarray}
G_{(5)}&=&dC_{(4)}+C_{(3)m}\wedge F^m+ \frac{1}{120}K_{mnpqt}A^m\wedge A^n\wedge A^p\wedge A^q\wedge A^t
\nonumber\\
&&+\frac{1}{2}\epsilon_{ab}B^a_{(2)}\wedge H^b_{(3)}- \frac{1}{6}\epsilon_{ab}M^a_{mnp}B^b_{(2)}\wedge A^m\wedge A^n\wedge A^p
\nonumber\\\nonumber\\
G_{(4)m}&=&DC_{(3)m}+C_{(2)mn}\wedge F^n+\frac{1}{24}K_{mnpqt}A^n\wedge A^p\wedge A^q\wedge A^t
\nonumber\\
&&+\frac{1}{2}\epsilon_{ab}B_{(2)}^a\wedge H^b_{(2)m}- \frac{1}{2}\epsilon_{ab}B_{(2)}^a\wedge H^b_{(3)}
\nonumber\\
&&+ \frac{1}{6}\epsilon_{ab}M^a_{qnp}B_{(1)m}^b\wedge A^n\wedge A^p\wedge A^q -\frac{1}{2}\epsilon_{ab} M^a_{mnp}B^b_{(2)}\wedge A^n\wedge A^p
\nonumber\\\nonumber\\
G_{(3)mn}&=&DC_{(2)mn}+C_{(2)p}f^p_{mn}+C_{(1)mnp}\wedge F^p+\frac{1}{6}K_{mnpqt}A^p\wedge A^q\wedge A^t
\nonumber\\
&&+\frac{1}{2}\epsilon_{ab} B^a_{(0)mn} H^b_{(3)} +\frac{1}{2}\epsilon_{ab}B^a_{(1)m}\wedge H^b_{(2)n}+\frac{1}{2}\epsilon_{ab}B^a_{(2)}\wedge
H^b_{(1)mn}
\nonumber\\
&&-\epsilon_{ab}\left( M^a_{qtp}B_{(0)mn}A^q\wedge A^t+M^a_{mpq}B^b_{(1)n}\wedge A^q+M_{mnp}^aB^b_{(2)} \right)\wedge A^p\nonumber\\
G_{(2)mnp}&=&DC_{(1)mnp}+C_{(0)mnpq}F^q+\frac{1}{2}K_{mnpqt}A^q\wedge A^t\nonumber\\
&&+\frac{3}{2}\epsilon_{ab} B^a_{(2)}B^b_{(0)[m|q}f^q_{|np]}+\frac{3}{2}\epsilon_{ab}B^a_{(0)mn}H^b_{(2)p}-
\frac{3}{2}\epsilon_{ab}B^a_{(1)m}\wedge H^b_{(1)np}
\nonumber\\
&&-\epsilon_{ab}\left( M_{mnp}^aB^b_{(2)}-3M^a_{mnq}B^b_{(1)p}\wedge A^q+\frac{3}{2}M^a_{pqt}B^b_{(0)mn}A^q\wedge A^t \right)
\end{eqnarray}
\begin{eqnarray}
G_{(1)mnpq}&=&DC_{(0)mnpq}+6C_{(1)pqt}f^t_{mn}+K_{mnpqt}A^t \nonumber\\&&+ 3\epsilon_{ab}B^a_{(0)mn}H^b_{(1)pq}
+6\epsilon_{ab}B^a_{(1)m}B^b_{(0)[n|t}f^t_{|pq]} \nonumber\\&& - \epsilon_{ab}\left( 6M^a_{mnt}B^b_{(0)pq}\wedge A^t - 4M^a_{mnp}B^b_{(1)q}
\right)
 \nonumber\\\nonumber\\
G_{(0)mnpqt}&=&-2C_{(0)spqt}f^s_{mn}+K_{mnpqt}
\nonumber\\
&&+30\epsilon_{ab}B^a_{(0)mn}B^b_{(0)[p|s}f^s_{|qt]} - 10\epsilon_{ab}M^a_{mnp}B^b_{(0)qt}
\end{eqnarray}
where the $G_R$ covariant derivatives are
\begin{equation}
D\psi_{(p)m_1m_2...m_q}=d\psi_{(p)m_1m_2...m_q}+(-)^p\psi_{(p)[m_1m_2...m_{q-1}|n}f_{|m_q]p}{}^n\wedge A^p
\end{equation}
The reduced Bianchi identities for the self-dual five form are
\begin{eqnarray}
dG_{(5)}-G_{(4)m}\wedge F^m&=&\frac{1}{2}\epsilon_{ab}{\cal H}^a_{(3)}\wedge {\cal H}^b_{(3)}\nonumber\\
DG_{(4)m}-G_{(3)mn}\wedge F^n&=&\epsilon_{ab}{\cal H}^a_{(3)}\wedge {\cal H}^b_{(2)m}\nonumber\\
DG_{(3)mn}-f_{mn}{}^pG_{(4)p}-G_{(2)mnp}\wedge F^p&=&\epsilon_{ab}\left({\cal H}^a_{(3)}\wedge {\cal H}^b_{(1)mn}+{\cal H}^a_{(2)m}\wedge
H^b_{(2)n}\right)\nonumber\\
DG_{(2)mnp}+O_{mnp}^{qt}G_{(3)qt}-G_{(1)mnpq}\wedge F^q&=&\epsilon_{ab}\left({\cal H}^a_{(3)}\wedge {\cal H}^b_{(0)mnp}-3{\cal H}^a_{(2)m}\wedge
H^b_{(2)np}\right)\nonumber\\
DG_{(1)mnpq}-O_{mnpq}^{tsl}G_{(2)tsl}-G_{(0)mnpqt}\wedge F^t&=&\epsilon_{ab}\left(4{\cal H}^a_{(2)[m} {\cal H}^b_{(0)npq]}+3{\cal
H}^a_{(1)[mn}\wedge
{\cal H}^b_{(0)pqt]}\right)\nonumber\\
DG_{(0)mnpqt}+O_{mnpqt}^{slij}G_{(1)slij}&=&10\epsilon_{ab}{\cal H}^a_{(1)[mn} {\cal H}^b_{(0)pqt]}\nonumber\\
O_{mnpqts}^{lijkh}G_{(0)lijkh}&=&10\epsilon_{ab}{\cal H}^a_{(0)[mnp} {\cal H}^b_{(0)qts]}
\end{eqnarray}
and for the three form
\begin{eqnarray}
d{\cal H}^a_{(3)}+{\cal H}^a_{(2)m}\wedge F^m&=&0\nonumber\\
D{\cal H}^a_{(2)m}+{\cal H}^a_{(1)mn}\wedge F^n&=&0\nonumber\\
D{\cal H}^a_{(1)mn}+{\cal H}^a_{(0)mnp} F^p&=&0\nonumber\\
D{\cal H}^a_{(0)mnp}&=&0\nonumber\\
\end{eqnarray}

\section{Right Action of the Group Manifold}

The identified group manifold $\cX=G/\G$ inherits the right action of the group $G_R$ on $G$. The calculation of how the reduced fields
transform under $G_R$ is somewhat involved and it is helpful to clarify the discussion by first considering the simpler case of the two form
transformation. The action of $G_R$ on the two-form is generated by the Lie derivative ${\cal L}_\omega$
\begin{equation}\label{lie derivative}
{\cal L}_\omega(\widehat{\cal B}^a_{(2)})={\cal L}_\omega(\widehat{B}^a_{(2)})+{\cal L}_\omega(\varpi^a_{(2)})=0
\end{equation}
where $\omega=\omega(x)$ is the parameter associated to the right action on the group manifold $G$. Using the fact that the Lie derivative may
be written ${\cal L}_\omega=\imath_{\omega}d+d\imath_{\omega}$, ${\cal L}_\omega(\nu^m)=-\nu^nf_{np}{}^m\omega^p$ and choosing the convention
$\imath_{\omega}\sigma^m=-\omega^m$, the transformations in (\ref{lie derivative}) imply
\begin{equation}
{\cal L}_{\omega}(\widehat{B}^a_{(2)})=-{\cal L}_\omega(\varpi^a_{(2)})=\frac{1}{2}M_{mnp}{}^a\omega^p\sigma^m\wedge\sigma^n+d\Xi^a_{(1)}
\end{equation}
where $\Xi^a_{(1)}\equiv \imath_{\omega}\varpi_{(2)}^a$. $\widehat{B}^a_{(2)}$ is in addition transformed by
$\delta(\Xi)_Y\widehat{B}^a_{(2)}=-d\Xi^a_{(1)}$ defining a gauge transformation $\delta_Z(\omega)$ which is independent of the internal
coordinates $y^i$
\begin{equation}
\delta_Z(\omega)\widehat{B}^a_{(2)}=\frac{1}{2}M_{mnp}{}^a\omega^p\sigma^m\wedge\sigma^n
\end{equation}
i.e.
\begin{equation}
\delta_Z(\omega)\widehat{\cal B}^a_{(2)}={\cal L}_\omega(\widehat{\cal B}^a_{(2)})-\delta_Y(\Xi)\widehat{\cal
B}^a_{(2)}=-\delta_Y(\Xi)\widehat{\cal B}^a_{(2)}
\end{equation}
where the second equality comes from the fact that the $\widehat{\cal B}^a_{(2)}$ is invariant under general coordinate transformations
generated by ${\cal L}_{\omega}$. The reduced components of $\widehat{\cal B}^a_{(2)}$ transform as
\begin{eqnarray}
\delta_Z(\omega)B^a_{(2)}&=&\frac{1}{2}M_{mnp}{}^a\omega^pA^m\wedge A^n\nonumber\\
\delta_Z(\omega)B^a_{(1)m}&=&B_{(1)n}^af_{mp}{}^n\omega^p-M_{mnp}{}^a\omega^pA^n\nonumber\\
\delta_Z(\omega)B^a_{(0)mn}&=&2B_{(1)[m|p}^af_{|n]q}{}^p\omega^q+M_{mnp}{}^a\omega^p
\end{eqnarray}
The corresponding transformation for the four form potential requires more care as the flux for this potential (\ref{four form flux def}) is
more complicated. The symmetry transformation of interest is that generated by $\delta_Z(\omega)$ rather than ${\cal L}_\omega$ and are defined
by ${\cal L}_\omega(G_{(5)})=0$, which requires the four form potential to transform as
\begin{eqnarray}\label{four form diffeo}
\delta_Z(\omega)\widehat{\cal C}_{(4)}&=&{\cal L}_{\omega}(\widehat{\cal C}_{(4)})-\delta_Y(\Xi)\widehat{\cal C}_{(4)}\nonumber\\
&=&-\delta_Y(\Xi)\widehat{\cal C}_{(4)}\nonumber\\
&=&\frac{1}{2}\epsilon_{ab}\Xi^a_{(1)}\wedge\widehat{\cal H}^b_{(3)}+d\chi_{(3)}
\end{eqnarray}
where the $\delta_Y(\Xi)$ transformation is only defined up to the arbitrary total derivative $d\chi_{(3)}$ and it must be stressed that
$\widehat{\cal C}_{(4)}$ is invariant under the diffeomorphism transformation ${\cal L}_{\omega}$, but not $\delta_Z(\omega)$. Combining
(\ref{four form flux def}) and (\ref{four form diffeo}), the symmetry transformation of interest is
\begin{equation}\label{C diffeomeorphism}
\delta_Z(\omega)\widehat{S}_{(4)}=\delta_Z(\omega)\left(\frac{1}{2}\varpi^a_{(2)}\wedge\widehat{B}^b_{(2)}
-\varpi_{(4)}\right)+\frac{1}{2}\epsilon_{ab}\Xi^a_{(1)}\wedge\widehat{\cal H}^b_{(3)}+d\chi_{(3)}
\end{equation}
It will be shown that, although ${\cal L}_{\omega}\widehat{S}$ is not $y$-independent, $\delta_Z(\omega)\widehat{S}$ is. The transformation
$\delta_Z(\omega)d\varpi_{(4)}={\cal L}_{\omega}(d\varpi_{(4)})$ will be calculated first\footnote{In all the variations of the fluxes
$\delta_Z(\omega)\varpi={\cal L}_{\omega}(\varpi)$ as the fluxes are invariant under the $\delta_Y$ and $\delta_X$ transformations.}. Consider,
\begin{eqnarray}\label{K diffeo}
{\cal L}_{\omega}({\cal K}_{(5)})&=&\left(\imath_{\omega}d+d\imath_{\omega}\right)\frac{1}{120}K_{mnpqt}
\sigma^m\wedge\sigma^n\wedge\sigma^p\wedge\sigma^q\wedge\sigma^t\nonumber\\
&=&-\frac{1}{24}K_{[mnpq|t}f_{|l]s}{}^t \omega^s\sigma^m\wedge\sigma^n\wedge\sigma^p\wedge\sigma^q\wedge\sigma^l\nonumber\\&&-
\frac{1}{24}K_{mnpqt} d\omega^t\wedge\sigma^m\wedge\sigma^n\wedge\sigma^p\wedge\sigma^q
\end{eqnarray}
and the transformation
\begin{equation}\label{2 flux diffeo}
{\cal L}_{\omega}(\varpi^a_{(2)})=-\frac{1}{2}M_{mnp}{}^a\omega^p\sigma^m\wedge\sigma^n+d\Xi_{(1)}^a
\end{equation}
Putting together (\ref{4 Flux}), (\ref{K diffeo}) and (\ref{2 flux diffeo}) the transformation of $d\varpi_{(4)}$ is therefore
\begin{eqnarray}\label{varpi_4 diffeo}
{\cal L}_{\omega}(d\varpi_{(4)})&=&{\cal L}_{\omega}\left(-\frac{1}{2}\epsilon_{ab}\varpi_{(2)}^a\wedge
d\varpi_{(2)}^b+{\cal K}_{(5)}\right)\nonumber\\
&=&-\frac{1}{12}\epsilon_{ab}\left(-\frac{1}{2}M_{[mn|p}{}^a\omega^p\sigma^m\wedge\sigma^n+d\Xi_{(1)}^a\right)\wedge
M_{|qts]}{}^b\sigma^q\wedge\sigma^t\wedge\sigma^s
\nonumber\\
&&+\frac{1}{4}\epsilon_{ab}\varpi_{(2)}^a\wedge M_{mnp}{}^bd(\omega^p\sigma^m\wedge\sigma^n)+{\cal L}_\omega({\cal K}_{(5)})
\end{eqnarray}
which can be written as
\begin{equation}\label{four flux variation 1}
{\cal L}_{\omega}(d\varpi_{(4)})=-d\theta_{(4)} -\frac{1}{2}\epsilon_{ab}d\varpi^a_{(2)}\wedge M_{mnp}{}^b\omega^p\sigma^m\wedge\sigma^n+{\cal
L}_{\omega}({\cal K}_{(5)})
\end{equation}
where
\begin{equation}
\theta_{(4)}=\frac{1}{2}\epsilon_{ab}\varpi^a_{(2)}\wedge\left(-\frac{1}{2}M_{mnp}{}^b\omega^p\sigma^m\wedge\sigma^n-d\Xi^b_{(1)}\right)
\end{equation}
Now consider the second term in the above expression (\ref{four flux variation 1})
\begin{eqnarray}\label{four flux variation 2}
\frac{1}{2}\epsilon_{ab}d\varpi^a_{(2)}\wedge M_{mnp}{}^b\omega^p\sigma^m\wedge\sigma^n=
\frac{1}{12}\epsilon_{ab}M_{[mnp}{}^aM_{qt]s}{}^b\omega^s\sigma^m\wedge\sigma^n\wedge\sigma^p\wedge\sigma^q\wedge\sigma^t
\end{eqnarray}
Using the fact that $\epsilon_{ab}M_{[mnp}{}^aM_{qt]s}{}^b=\epsilon_{ab}M_{[mnp}{}^aM_{qts]}{}^b$, (\ref{four flux variation 2}) may be written
as
\begin{equation}\label{four flux variation 3}
\frac{1}{12}\epsilon_{ab}M_{[mnp}{}^aM_{qts]}{}^b\omega^s\sigma^m\wedge\sigma^n\wedge\sigma^p\wedge\sigma^q\wedge\sigma^t
\end{equation}
and now making use of the identity $2\epsilon_{ab}M_{[mnp}{}^aM_{qts]}{}^b+3K_{[mnpq|l}f_{|ts]}{}^l=0$ (\ref{flux constraints}) to write
(\ref{four flux variation 3}) as
\begin{eqnarray}
\frac{1}{2}\epsilon_{ab}d\varpi^a_{(2)}\wedge M_{mnp}{}^b\omega^p\sigma^m\wedge\sigma^n&=&
-\frac{1}{24}K_{[mnpq|l}f_{|ts]}{}^l\omega^s\sigma^m\wedge\sigma^n\wedge\sigma^p\wedge\sigma^q\wedge\sigma^t\nonumber\\&&
+\frac{1}{12}K_{[mnp|sl}f_{|tq]}{}^l\omega^s\sigma^m\wedge\sigma^n\wedge\sigma^p\wedge\sigma^q\wedge\sigma^t\nonumber\\
\end{eqnarray}
and substituting it into (\ref{varpi_4 diffeo}), the variation of $d\varpi_{(4)}$ becomes
\begin{eqnarray}\label{variation}
{\cal L}_{\omega}(d\varpi_{(4)})&=&-d\theta_{(4)} -\frac{1}{2}\epsilon_{ab}d\varpi^a_{(2)}\wedge M_{mnp}{}^b\omega^p\sigma^m\wedge\sigma^n+{\cal
L}_{\omega}({\cal
K}_{(5)})\nonumber\\&=&-d\theta_{(4)}-\frac{1}{24}K_{mnpqt}d\omega^t\wedge\sigma^m\wedge\sigma^n\wedge\sigma^p\wedge\sigma^q\nonumber\\
&&+\frac{1}{12}K_{[mnp|qt}f_{|sl]}{}^t\omega^q\wedge\sigma^m\wedge\sigma^n\wedge\sigma^p\wedge\sigma^s\wedge\sigma^l\nonumber\\&=&
d\left(-\theta_{(4)}-\frac{1}{24}K_{mnpqt}\omega^t\sigma^m\wedge\sigma^n\wedge\sigma^p\wedge\sigma^q\right)
\end{eqnarray}
It is simple to show that, acting on the space of forms, $[{\cal L}_{\omega},d]=0$ and therefore the variation of $\varpi_{(4)}$ commutes with
the total derivative so that the expression (\ref{variation}) can be integrated to give
\begin{eqnarray}
{\cal L}_{\omega}(\varpi_{(4)})&=&-\theta_{(4)}-\frac{1}{24}K_{mnpqt}\omega^t\sigma^m\wedge\sigma^n\wedge\sigma^p\wedge\sigma^q+d\Omega_{(3)}\nonumber\\
&=&-\frac{1}{2}\epsilon_{ab}\varpi^a_{(2)}\wedge\left(-\frac{1}{2}M^b_{mnp}\omega^p\sigma^m\wedge\sigma^n-d\Xi^b_{(1)}\right)
\nonumber\\&&-\frac{1}{24}K_{mnpqt}\omega^t\sigma^m\wedge\sigma^n\wedge\sigma^p\wedge\sigma^q+d\Omega_{(3)}\nonumber\\
\end{eqnarray}
for some arbitrary three form $\Omega_{(3)}$. Now that ${\cal L}_{\omega}(\varpi_{(4)})$ has been determined, the variation of the second term
in (\ref{C diffeomeorphism}) is considered
\begin{eqnarray}
\delta_Z({\omega})\left(-\frac{1}{2}\epsilon_{ab}\varpi^a_{(2)}\wedge \widehat{B}^b_{(2)}\right)&=&-\frac{1}{2}\epsilon_{ab}{\cal
L}_{\omega}(\varpi^a_{(2)})\wedge \widehat{B}^b_{(2)}-\frac{1}{2}\epsilon_{ab}\varpi^a_{(2)}\wedge
\left(\delta_Z(\omega)\widehat{B}^b_{(2)}\right)
\nonumber\\&=&-\frac{1}{2}\epsilon_{ab}\left(-\frac{1}{2}M_{mnp}{}^a\omega^p\sigma^m\wedge\sigma^n +d\Xi^a_{(1)}\right)\wedge
\widehat{B}^b_{(2)} \nonumber\\&&- \frac{1}{4}\epsilon_{ab}\varpi^a_{(2)}\wedge M_{mnp}{}^b\omega^p\sigma^m\wedge\sigma^n
\end{eqnarray}
where a gauge transformation is incorporated into $\delta(\omega)\widehat{B}^b_{(2)}$ to give $\delta_Z(\omega)\widehat{B}^b_{(2)}$. Putting
these results together gives
\begin{eqnarray}\label{junk diffeo}
\delta_Z({\omega})\left(-\frac{1}{2}\epsilon_{ab}\varpi^a_{(2)}\wedge\widehat{B}^b_{(2)}+\varpi_{(4)}\right)
&=&\frac{1}{4}\epsilon_{ab}M_{mnp}{}^a\omega^p\sigma^m\wedge\sigma^n\wedge\widehat{B}^b_{(2)}
\nonumber\\
&&-\frac{1}{2}\epsilon_{ab}d\Xi^a_{(1)}\wedge\widehat{B}^b_{(2)}+\frac{1}{2}\varpi^a_{(2)}\wedge d\Xi^b_{(2)}+d\Omega_{(3)}
\nonumber\\
&&-\frac{1}{24}K_{mnpqt}\omega^t\sigma^m\wedge\sigma^n\wedge\sigma^p\wedge\sigma^q
\end{eqnarray}
The first two terms in the last line can be written as
\begin{equation}\label{junk diffeo 2}
\frac{1}{2}\epsilon_{ab}d\Xi^a_{(1)}\wedge\widehat{B}^b_{(2)}-\frac{1}{2}\varpi^a_{(2)}\wedge
d\Xi^b_{(2)}=d\left(\frac{1}{2}\epsilon_{ab}\Xi^a_{(1)}\wedge{\widehat{\cal B}}^b_{(2)}\right)+\frac{1}{2}\epsilon_{ab}\Xi^a_{(1)}\wedge{\cal
H}^b_{(3)}
\end{equation}
and the first term in (\ref{junk diffeo 2}) can be removed by choosing an appropriate value for $\Omega_{(3)}$ in (\ref{junk diffeo}) and the
last term of (\ref{junk diffeo 2}) cancels with the $\delta_Y(\Xi)$ transformation of (\ref{C diffeomeorphism}) to leave the transformation
\begin{equation}\label{transformation}
\delta_Z(\omega)\widehat{S}_{(4)}=\frac{1}{4}\epsilon_{ab}M_{mnp}{}^a\omega^p\sigma^m\wedge\sigma^n\wedge\widehat{B}^b_{(2)}
-\frac{1}{24}K_{mnpqt}\omega^t\sigma^m\wedge\sigma^n\wedge\sigma^p\wedge\sigma^q
\end{equation}
Substituting the Scherk-Schwarz ansatz for the potentials (\ref{C def}) and (\ref{B def}) into (\ref{transformation}), the gauge transformations
of the potentials are found to be
\begin{eqnarray}\label{C diffeomorphism 2}
\delta_Z(\omega^m)S_{(4)}&=&\frac{1}{4}\epsilon_{ab}M_{mnp}{}^a\omega^pB_{(2)}^b\wedge A^m\wedge A^n-\frac{1}{24}K_{mnpqt}\omega^tA^m\wedge
A^n\wedge A^p\wedge A^q
\nonumber\\\nonumber\\
\delta_Z(\omega^m)S_{(3)m}&=&S_{(3)n}f_{mp}{}^n\omega^p+\frac{1}{4}\epsilon_{ab}M_{npq}{}^a\omega^qA^n\wedge A^p\wedge
B^b_{(1)m}\nonumber\\
&&+\frac{1}{2}\epsilon_{ab}M_{mnp}{}^a\omega^pA^n\wedge B^b_{(2)} +\frac{1}{6}K_{mnpqt}\omega^tA^n\wedge A^p\wedge A^q
\nonumber\\\nonumber\\
\delta_Z(\omega^m)S_{(2)mn}&=&2S_{(2)[m|p}f_{|n]q}{}^p\omega^q+\frac{1}{4}\epsilon_{ab}M_{pqt}{}^a\omega^tA^p\wedge
A^qB_{(0)mn}^b+\epsilon_{ab}M_{mpq}{}^a\omega^pA^q\wedge
B^b_{(1)n}\nonumber\\
&&+\frac{1}{2}\epsilon_{ab}M_{mnp}{}^a\omega^pB^b_{(2)} -\frac{1}{2}K_{mnpqt}\omega^tA^p\wedge A^q
\nonumber\\\nonumber\\
\delta_Z(\omega^m)S_{(1)mnp}&=&3S_{(1)[mn|q}f_{|p]t}{}^q\omega^t+ \frac{3}{2}\epsilon_{ab}M_{mnq}{}^a\omega^q B^b_{(1)p}
-\frac{3}{2}\epsilon_{ab}M_{mqt}{}^a\omega^t B^b_{(0)np}A^q\nonumber\\
&&+K_{mnpqt}\omega^tA^q
\nonumber\\\nonumber\\
\delta_Z(\omega^m)S_{(0)mnpq}&=&4S_{(0)[mnp|t}f_{|q]s}{}^t\omega^s+3\epsilon_{ab}M_{[mn|t}{}^aB^b_{(0)|pq]}\omega^t-K_{mnpqt}\omega^t
\end{eqnarray}

\end{document}